# Privacy Preservation for Wireless Sensor Networks in Healthcare: State of the Art, and Open Research Challenges

Yasmine N. M. Saleh [a,*], Claude C. Chibelushi [b], Ayman A. Abdel-Hamid [a], and Abdel-Hamid Soliman [c]

[a] College of Computing and Information Technology, Arab Academy for Science and Technology and Maritime Transport, Alexandria, Egypt.
[b] School of Computing, Staffordshire University, Beaconside, Stafford ST18 0AD, U.K.
[c] School of Engineering, Staffordshire University, Mellor building, College Rd, Stoke on Trent, ST4 TDE, U.K.

**ARTICLE INFO**

**ABSTRACT**

The advent of miniature biosensors has generated numerous opportunities for deploying wireless sensor networks in healthcare. However, an important barrier is that acceptance by healthcare stakeholders is influenced by the effectiveness of privacy safeguards for personal and intimate information which is collected and transmitted over the air, within and beyond these networks. In particular, these networks are progressing beyond traditional sensors, towards also using multimedia sensors, which raise further privacy concerns. Paradoxically, less research has addressed privacy protection, compared to security. Nevertheless, privacy protection has gradually evolved from being assumed an implicit by-product of security measures, and it is maturing into a research concern in its own right. However, further technical and socio-technical advances are needed. As a contribution towards galvanising further research, the hallmarks of this paper include: (i) a literature survey explicitly anchored on privacy preservation, it is underpinned by untangling privacy goals from security goals, to avoid mixing privacy and security concerns, as is often the case in other papers; (ii) a critical survey of privacy preservation services for wireless sensor networks in healthcare, including threat analysis and assessment methodologies; it also offers classification trees for the multifaceted challenge of privacy protection in healthcare, and for privacy threats, attacks and countermeasures; (iii) a discussion of technical advances complemented by reflection over the implications of regulatory frameworks; (iv) a discussion of open research challenges, leading onto offers of directions for future research towards unlocking the door onto privacy protection which is appropriate for healthcare in the twenty-first century.

This work was supported in part by the Arab Academy for Science, Technology and Maritime Transport (Egypt) under its funding scheme for doctoral research, with a doctoral study grant awarded to Y. N. M. Saleh. The work of C. C. Chibelushi was supported in part by St. John's College (University of Oxford, U.K.) through a Visiting Scholarship.

Yasmine N. M. Saleh is with the College of Computing and Information Technology, Arab Academy for Science and Technology and Maritime Transport, Alexandria, Egypt. (e-mail: yasmine_nagi@ aast.edu).

C. C. Chibelushi is with the School of Computing, Staffordshire University, Beaconside, Stafford ST18 0AD, U.K. (e-mail: C.C.Chibelushi@staffs.ac.uk).

Ayman A. Abdel-Hamid is with the College of Computing and Information Technology, Arab Academy for Science and Technology and Maritime Transport, Alexandria, Egypt. (e-mail: hamid@ aast.edu).

Abdel-Hamid Soliman is with the School of Engineering, Staffordshire University, Mellor building, College Rd, Stoke on Trent, ST4 TDE, U.K. (e-mail: A.Soliman@staffs.ac.uk).

## 1 INTRODUCTION

THE advent of miniature wearable or implantable biosensors, together with other wireless sensors, coupled to significant advances in data processing and wireless communication techniques, have generated many opportunities for the deployment of wireless sensor networks in healthcare. However, ensuring the privacy of personal health information is paramount in healthcare systems, hence research into appropriate privacy protection for such networks is an important item on the research agenda for the twenty-first century.

This section introduces wireless sensor networks and discusses the relative neglect of privacy protection (in comparison to security) in the research literature. In addition, this section delineates scope of the paper, and reviews the content of related literature surveys which cover privacy in wireless sensor networks in general or for healthcare in particular. The section also highlights what makes this paper significantly different from the other surveys and summarizes the contributions of the paper.

### 1.1 Wireless sensor networks

A wireless sensor network (WSN) is a self-organizing multi-hop wireless network of nodes, which can be made up of tens to thousands of (typically) very small low-power sensor devices. These devices are deployed to sense the relevant phenomena, such as heat, motion, light, sound, air quality, heart or brain activity, blood pressure or oxygen saturation in the human body. Generally, the sensor devices collect data (from the surrounding environment or the person to whom they are attached, for example), then they wirelessly send the data to collection devices called base stations [1]. A base station can serve as gateway to another network (such as the Internet) which has a wider spatial coverage than the wireless sensor network.

Possible areas of applications of wireless sensor networks are: healthcare; industrial automation; metropolitan, military or environmental monitoring; animal tracking; civil engineering; logistics and transportation; and sports [2]. It was estimated that the largest impact of the Internet of Things — which is a broader technology domain encompassing wireless sensor networks — will be in healthcare, with a projected economic impact of $1.1 trillion to $2.5 trillion per year by 2025; among the predictions, it was forecast that remote monitoring could reduce treatment costs for chronic diseases by 10 to 20 percent [3].

Physical health and mental health are important ingredients of human life. They make a major contribution to the quality of life and to the economic development of individuals, communities and countries. The provision of quality healthcare, to treat or (ideally) prevent health problems, is thus acknowledged as a worthy priority in modern society. As sensors and wireless technologies have developed up to sufficient maturity for healthcare applications, wireless sensor networks are increasingly used in the context of healthcare. Some application scenarios and benefits accruing from wireless sensor networks are presented in later sections of this paper.

This survey paper targets healthcare as the selected application domain for wireless sensor networks, with focus on privacy preservation. The choice has been driven by the importance of quality healthcare, coupled to the promises of wireless sensor networks in healthcare, and the paramount importance of preserving the privacy of healthcare subjects and stakeholders [3], which is a challenge that requires more research efforts dedicated to it.

### 1.2 Privacy preservation in wireless sensor networks for healthcare: Another Cinderella?

A look at developments over the years reveals that the challenge of enhancing the security of wireless sensor networks has received far more attention, from researchers and practitioners, than privacy protection. The latter is often tackled as an implicit by-product of security. With particular reference to healthcare, although the use of wireless sensor networks in the healthcare sector has been investigated by researchers, comparatively little effort has been dedicated to privacy protection [4]. A significant amount of research on WSN-based healthcare systems has been devoted to the physical design of the network, for example, focusing on criteria such as system reliability, power consumption and cost effectiveness. Although the importance of privacy and security is well-acknowledged for sensitive personal data, like data about the health of individuals, a considerable research effort is still needed to develop robust solutions for wireless sensor networks used in the context of healthcare.

Despite technological advances in wireless sensor networks for healthcare applications, which have enhanced the feasibility of continuously monitoring patients or healthy individuals, it is no secret that the successful adoption of WSN-based healthcare systems will also depend directly on the privacy and security which these systems would be able to provide [5]. Widespread adoption will also be dictated by how the legal implications of privacy are dealt with; for example, with regards to the use and ownership of data collected by wireless sensor networks. Privacy and security violations may lead to a leakage of sensitive information about the diseases of patients, which may be embarrassing or critical, and which could cause the patients to lose their employment or be unable to obtain insurance, and sometimes lead to risks such as an adversary (possibly a criminal mind) finding the location of a person, with potentially life-threatening consequences [5]. Consequently, effective measures against both data content and contextual privacy violations are an indispensable prerequisite in most WSN-based applications, and they are of paramount importance in healthcare applications [6].

### 1.3 Scope

Sadly, the word 'privacy' is often used fuzzily in the literature. In particular, there is a general tendency in the literature to mix aspects where privacy protection is distinguishable from security, with aspects which fall in the common ground between the privacy goals of a system and its security goals. It is therefore important to adopt clear definitions so as to delineate the scope of for this paper.

Firstly, the definitions of privacy and security adopted in this survey are summarized below; more details are given in the appendix. Information privacy is the *right of an individual to control his or her identifiable data*, with regards to acquisition, use, or disclosure. On the other hand, the central tenet of information security is the *protection of information and information systems*, with regards to unauthorized access, use, disclosure, disruption, modification, or destruction. Thus, as both information privacy and security relate to the *protection or control of information or data*, security mechanisms can contribute to the attainment of information privacy but they are not sufficient on their own. An important consideration is that the definition of security does not make explicit reference to control by the *individual*. Hence, other protection mechanisms, in addition to security mechanisms, are required to cover aspects related to the individual (also known as data subject).

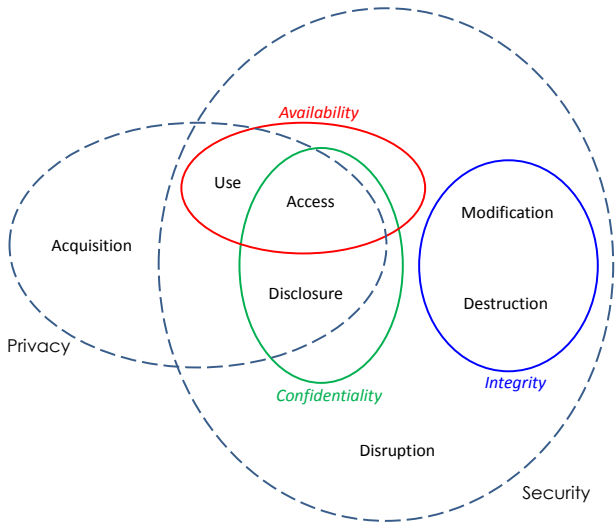

Figure 1. Overlap between keywords included in the definitions of health information privacy and security (detailed definitions are given in the appendix)

One can see from the Venn diagram given in Figure 1 that the areas of explicit overlap between information privacy and information security are around *controlled information disclosure, use and access*. With reference to the so-called "CIA triad" (confidentiality, integrity, availability), privacy and security both vie for preserving confidentiality, but privacy places on information availability some restrictions which are dictated by the individual whom the information is about. With regards to integrity (the third goal of the triad), it should be remembered that breaches of information and system integrity can constitute or lead to a violation of information privacy.

The literature often jumbles up privacy concerns and conventional security concerns. This paper *focuses on privacy protection which is not provided by mechanisms from the realm of conventional security concerns*. The survey presented herein untangles privacy and security concerns by decoupling privacy goals from security goals. The privacy goals then provide the guiding thread for the review of the literature on privacy protection for wireless sensor networks, with a focus on privacy services and their underpinning mechanisms to achieve privacy goals. The authors of this paper acknowledge the relevance and importance of security goals, but they refer the interested reader to the relevant literature, where applicable.

The second set of important definitions relates to the distinction between *hard privacy* and *soft privacy*, as discussed in Deng et al. [16], who quoted the work of Danezis. The data protection goal of hard privacy is achieved through *data minimization*, whereby a data subject divulges as little personal data as possible to third parties, thus attempting to reduce the need to trust others. On the other hand, soft privacy starts from the premise that the data subject lost control of personal data. The data subject thus relies on trust in the honesty and competence of the data controller; the latter is responsible for the data. Hence, the data protection goal of soft privacy is achieved through policies, access control, and audit; to provide data security and to process data for a specific purpose and with consent.

This paper *focuses primarily on hard privacy*. A brief discussion of soft privacy issues is included, to acknowledge the technological implications (for privacy protection) which arise from privacy policies and regulatory frameworks. The main thrust of the paper is on privacy protection software, which is anchored on privacy goals.

Consequently, the third set of important definitions relates to the distinction between privacy goals and their security counterparts. The terminology adopted for privacy goals in this paper is the one proposed by Pfitzmann and Hansen [17]. The choice is guided by the fact that this terminology evolved from many years of consultation with the community of researchers and practitioners in the technical field of privacy protection. This paper is anchored on goals specifically defined for privacy; namely: *anonymity*, *pseudonymity*, *unlinkability*, *unobservability* and *undetectability* [17].The paper considers privacy protection as the attainment of privacy goals. The paper thus reviews privacy attacks and countermeasures by relating them (where possible) to the privacy goals which the attack violates or the countermeasures thrives to meet.

Although classical goals of information security, such as the CIA triad [16], are required to underpin some facets of privacy protection, the authors of this paper deem that information security has received such an extensive coverage in the literature, that security goals should be left out of the scope of the paper, to avoid making the paper inordinately long. Readers interested in security goals, and in the corresponding attacks, countermeasures and issues, are referred to the relevant literature which has a rich list of publications such as [18] [19] [20] [21] [22] [23] [24] [7] [25] [5].

Furthermore, acknowledging the multifaceted challenge of privacy preservation in healthcare applications (the complexity is discussed in Section 2.4.2, this paper focuses on privacy preservation techniques centered on the *healthcare data subject* (such as a patient or elderly). Thus, also out of scope of the paper are the regulatory and technical infrastructures which would take into account all key stakeholders (like the regulatory and enforcement authorities, healthcare agencies, …, in addition to the healthcare data subjects), when managing end-to-end privacy concerns in applications of wireless sensor networks in healthcare. The larger system within which a wireless sensor network would typically operate (for instance, computer networks which link health service providers such as staff, hospitals and other healthcare providers, and insurers) is thus also out of scope, and so is the privacy of stakeholders (such as doctors or lifestyle coaches) other than the healthcare data subject. The scope of the paper, in relation to privacy stakeholders, is summarized in Figure 2.

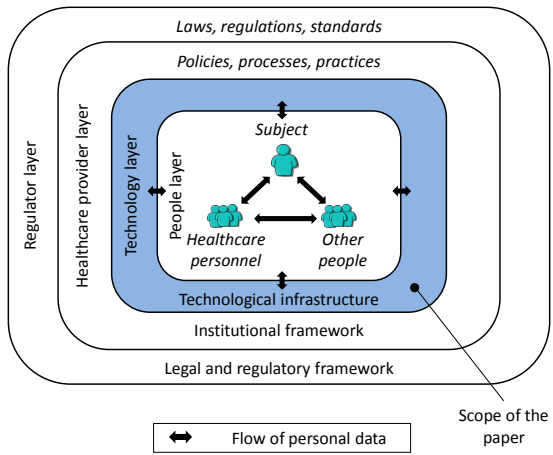

Figure 2. Scope of the paper in relation to privacy stakeholders

## 1.4  Related literature surveys

Several survey papers which primarily focus on privacy for wireless sensor networks have been published. Ordered in terms of the number of citations received (at the time of writing this paper), the most cited are: [7], [8], [26], [9], [27], [10], [11], [12], [6], [28] and [13]. As depicted in Figure 3, the most popular survey papers are [7], [8], [26] and [9]. Among the survey papers, those which tackle security and privacy, with focus on the healthcare application domain, are [7], [8], [9], [10], [11], [12] and [13].

The most popular survey paper [7] discusses the security and privacy requirements related to the storage and transmission of data in Wireless Body Area Networks (WBAN). The paper [7] views data privacy as an access control problem, whereby only authorized people should be able to access, view and use the patient-related data. The coverage of the paper is predominantly around security requirements; the paper views privacy only from the limited vantage point of access control, and it does not explicitly address privacy goals (advocated by the terminology proposed by Pfitzmann and Hansen [17].

The next most popular paper ([8]), discusses the challenge of achieving a balance between the security and privacy design goals for implanted medical devices (such as pacemakers, drug delivery systems and neuro-stimulators) and the effectiveness of the treatment and medical safety. The paper defines the main dimensions of privacy as: device-existence privacy, device-type privacy, specific-device ID privacy, measurement and log privacy and bearer privacy. The paper analyses the security and privacy goals for implanted medical devices, and the impact of achieving these goals, such as the tension between security goals and accessibility, device resources and usability. However, the paper does not cover the technical means (such as algorithms and specific techniques) to achieve the goals discussed in the paper.

[26] is the only paper among the most cited papers which focuses on privacy preservation techniques for wireless sensor networks in general (not focusing on healthcare, but with potential applicability to healthcare). The paper categorizes privacy concerns (and by implication, privacy protection techniques) in wireless sensor networks as either data privacy or context privacy. Data privacy is concerned with the privacy of the data collection and the queries issued in a wireless sensor network. Context privacy relates to the contextual information, such as the spatio-temporal context, hence leading to the concepts of location privacy and temporal privacy. The paper presents a taxonomy of privacy protection techniques. The rest of the paper is dedicated to discussing the techniques related to the privacy categories identified in the taxonomy. The paper concludes with a comparison between the techniques, based on

the level of privacy, accuracy, delay time and power consumption. Unlike the other two survey papers ([7] [8]), this paper [26] provides a coverage of privacy preservation techniques. However, the paper failed to mention privacy goals, such as unobservability, undetectability, unlinkability and pseudonymity, which were advocated by Pfitzmann and Hansen [17]. Furthermore, its coverage of privacy protection techniques is dominated by location privacy.

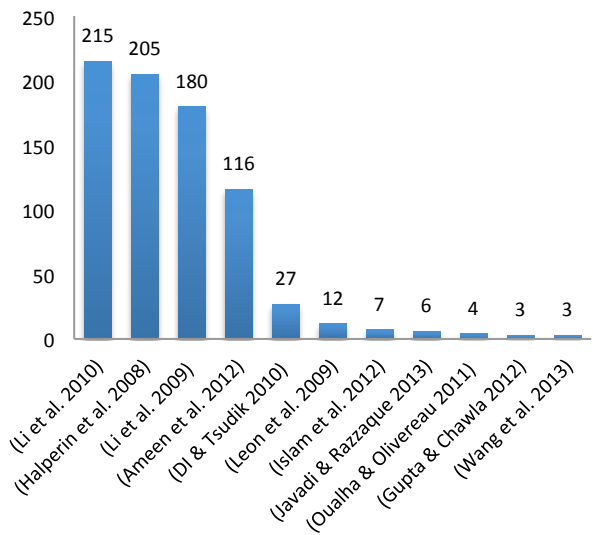

Figure 3. Numbers of citations of papers on privacy in wireless sensor networks (as at June 2015)

Modelled after [26], [28] adopts the same categorization of the privacy preservation techniques. It even presents a similar comparison as a conclusion for the paper. In addition, many of the techniques discussed in the paper were already covered in [26].

Another survey paper [9] is concerned with security and privacy issues in WSN-based healthcare systems. The paper presents a summary of selected WSN-based healthcare projects, security threats, attacks and counter measures. The paper views the privacy problem as one relating to where the data should be stored, who should access that data and who should be responsible for the maintenance of data. Their suggested methods for preserving privacy are: public awareness of privacy issues and solutions; user identification based on need; and encryption of communication in the wireless sensor network. The paper mainly focuses on wireless sensor network security and lacks a technical discussion of privacy preservation techniques.

Another survey paper [27] discusses security and privacy in wireless sensor networks. The paper includes three main subsections: wireless sensor networks, vehicular ad-hoc networks and disruption-tolerant networks. The paper discusses security and privacy issues in each of these three categories. However, the paper views privacy as a by-product of security; privacy is not covered in its own right.

The survey presented in [10] describes both the underlying technologies for wireless sensor networks and a review of chosen security-related and privacy-related work, in general and in healthcare applications in particular. However, its coverage of privacy protection is limited.

Following [7], the paper by [12] is concerned with the security and privacy issues for WBANs. In addition, both papers have the same view of data privacy as being an access control problem where only authorized personnel should be able to access patient-related information. The paper [12] views privacy as a by-product of security; mainly relying on security-related solutions (for data confidentiality, data access control, accountability, revocability, non-repudiation, policy requirement and public awareness) to ensure the privacy of the data. The paper clearly lacks focus on privacy solutions.

In [13], the focus is on security and privacy issues for patient related data in WSN-based healthcare systems. The paper views privacy requirements in e-health as anonymity and unlinkability requirements. It does not provide any details of actual techniques, to achieve these requirements, and ensure privacy.

The review of the previous survey papers has revealed that there is a lack of survey papers which primarily focus on issues related to privacy protection in WSN-based healthcare systems. None of the above survey papers adequately cover or discuss:

- The fulfilment of privacy goals from the perspective of WSN-based healthcare systems.
- An assessment of early work on WSN-based healthcare systems, based on ability to withstand privacy attacks.
- Privacy protection techniques which could be used in healthcare systems that are based on wireless sensor networks.
- Privacy threat analysis methodologies which can be deployed to identify the privacy services required by a WSN-based healthcare system.
- Assessment methodologies to gauge the vulnerabilities of an existing system or prospective system, and determine the level of privacy protection offered by the system.

### 1.5 Contributions of the paper

Given the importance of healthcare in the twenty-first century, coupled to the wide-ranging opportunities offered by wireless sensor networks and the crucial need for privacy protection, which has hitherto not been met adequately in the healthcare context, there is a need for a significant research effort focused on privacy protection techniques in wireless sensor networks for healthcare.

Some previously published survey papers discussed privacy for wireless sensor networks in general, and some focused on networks meant for use in healthcare systems in particular, as in [7], [8], [9], [10], [11], [12], [13], [14]. One of the differences between this survey and the others is that *it specifically focuses on privacy technology*, unlike the other papers which considered privacy and security together, and some even viewed privacy solely as an implicit by-product of security techniques. Furthermore, although some of the other survey papers highlighted the critical importance of privacy preservation in wireless sensor networks, some presented a rather limited discussion of the possible defenses, with no or little details given about techniques or methodologies, as in [15] and [10]. This survey emphasizes the technical aspect. To the best of our knowledge, there has not yet been a comprehensive survey paper which *untangles privacy goals from security goals*, and discusses *privacy preservation for wireless sensor networks in healthcare*, including a *review of threat analysis and assessment methodologies* for identifying privacy protection services required in a wireless sensor network for healthcare, or for gauging the vulnerabilities of such a system. These are some hallmarks of this paper.

To further contribute to the understanding of the state of the art, and to the formulation of future research on privacy protection, which is urgently needed, this paper also combines a *critical survey of the relevant technological developments with a reflection over the implications arising from the regulatory frameworks* within which the technology will operate. This combination is justified based on the fact that laws and policies will have a significant bearing on the operational requirements that privacy-preserving technology for wireless sensor networks will have to meet, in order to be ready for deployment in the real-world.

The primary aim of this paper is to survey the literature on the privacy of wireless sensor networks, which can be integrated into healthcare systems, and to fill the knowledge gaps left by previous survey papers. Thus, the objectives of this paper are to:

- Build upon the distinction between privacy goals and security goals, in order to focus on privacy and thus contribute to enhanced clarity and coherence of research into privacy preservation.
- Discuss the major privacy goals from the perspective of a healthcare system which is based on wireless sensor networks, and use these goals as anchor point for the discussion of such systems, with regards to resilience to privacy attacks.
- Present a survey of privacy protection techniques reported in the literature, which could be used in healthcare systems that are based on wireless sensor networks, or even wireless multimedia sensor networks.

- Present a review of threat analysis and assessment methodologies which could be deployed to identify privacy protection services required in wireless sensor networks for healthcare, or to gauge the vulnerabilities of an existing system or prospective system, and determine its level of privacy protection.
- Reflect over the implications, for technological developments in privacy protection, which arise from the regulatory frameworks within which the technology will operate.

The reader should be mindful that privacy 'goals' are interchangeably referred to in the literature as privacy 'properties', 'attributes', or 'basic building blocks'.

### 1.6 Structure of the paper

This section has introduced wireless sensor networks, discussed the relative neglect of privacy protection (in comparison to security) in the research literature, and summarized the contributions of the paper. In addition, this section summarizes the content of related literature surveys, and it highlights what makes this paper significantly different from the other surveys. The rest of the paper is organized as follows. Section 2 discusses healthcare applications of wireless sensor networks, broadly considered and also placing a special emphasis on wireless body sensor networks. The section also includes a discussion of privacy issues and their complexity in wireless sensor networks for healthcare. Section 3 reviews privacy services in wireless sensor networks. Section 4 then reviews developments towards healthcare systems, which are built on wireless multimedia sensor networks. Section 5 reviews privacy threat analysis methodologies and measurements for privacy assessment, with a summary of privacy attacks and countermeasures included as a prelude to the review of the methodologies and measurements. Section 6 discusses the regulatory frameworks for information privacy in healthcare and their implications for privacy protection technology. To stimulate future research towards turning privacy preservation in WSN-based healthcare systems into reality, section 7 discusses open research challenges, and offers directions for future research. The paper offers concluding remarks in Section 8. The appendices include a list of definitions and a list of acronyms.

## 2 HEALTHCARE APPLICATIONS OF WIRELESS SENSOR NETWORKS, AND THEIR PRIVACY IMPLICATIONS

Sensors deployed in a WSN-based healthcare system can be medical body sensors or other sensors such as those installed in the surrounding environment. Hence, this section includes sub-sections which group healthcare applications into those for wireless sensor networks in general and those for wireless body sensor networks (WBSNs) in particular. The section also includes a review of significant developments in wireless sensor networks, and of the general architecture of such networks in healthcare. Furthermore, the section discusses privacy issues and the imperative need for privacy protection in healthcare, and it spells out the multifaceted challenge of privacy protection in wireless sensor networks for healthcare.

## 2.1 Significant developments in wireless sensor networks

Spurred in the 1980s by the Defense Advanced Research Projects Agency (DARPA), of the United States, early research into wireless sensor networks was conducted through the Distributed Sensor Networks (DSN) program; it was targeted at the military [29][30]. A second wave of research was triggered by DARPA, through the Sensor Information Technology (SensIT) research program, at the turn of the twenty-first century. The turn of the century was also a significant milestone for standardization efforts, with the advent of standards such as IEEE 802.15.4 (Zigbee (network security)) and IP v6 (IETF (RFC 2460 and others), ITU M2M (with an initial focus on the health sector)).

Increased interests in wireless sensor networks was spurred at the dawn of the twenty-first century by the availability of inexpensive low-power miniature integrated devices, namely 'system on a chip' (SoC) — which combine sensor, processor, memory and radio communication — and also by micro-electromechanical systems (MEMS) and advances in wireless communication [29] [30]. Wireless sensor networks are now recognized as a very important technology for the twenty-first century [30]. According to [31], shipments of WSN chipsets will exceed 1 billion by 2017 (Figure 4); and IEEE 802.15.4 and ZigBee make up the largest market share.

Initial applications of wireless sensor networks were in the military area. Non-military uses saw the day around 1990; first in the industrial sector (factory automation, oil and gas, power transmission, environmental monitoring …) and later in the consumer sector (e-health, smart home, transportation …) (Figure 5). According to estimates reported in [3], the Internet of Things (which subsumes wireless sensor networks) will have its largest impact in healthcare. For example, remote monitoring has been forecast to yield a reduction of treatment costs for chronic diseases by 10 to 20 percent [3].

## 2.2 Applications of wireless sensor networks at large

In healthcare applications, wireless sensor nodes can be deployed inside or on the human body, or in the environment surrounding the person. They can be used to monitor the functions and health-related characteristics of the body or the behavior of the person. They can also enable the monitoring of the environment surrounding the person such as by measuring air quality or ambient temperature, light or noise. Wireless sensor networks have found uses in a wide diversity of areas, and they have opened new market opportunities [32]. The application of wireless sensor networks in the healthcare sector promises a significant improvement of the quality of care for a variety of segments of the human population [33]. Examples of possible applications are: health monitoring in mass casualty disasters and of patients in hospitals; assistive-living aids for patients or other people suffering from declining mental, sensory or motor capabilities; and medical and behavioral monitoring of the elderly.

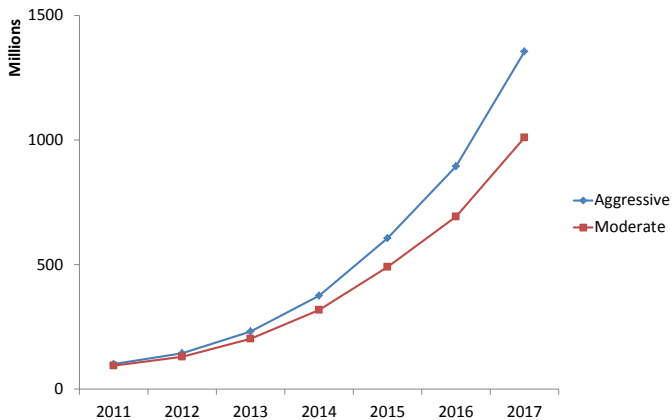

Figure 4. Global shipments of chipsets for wireless sensor networks [31]

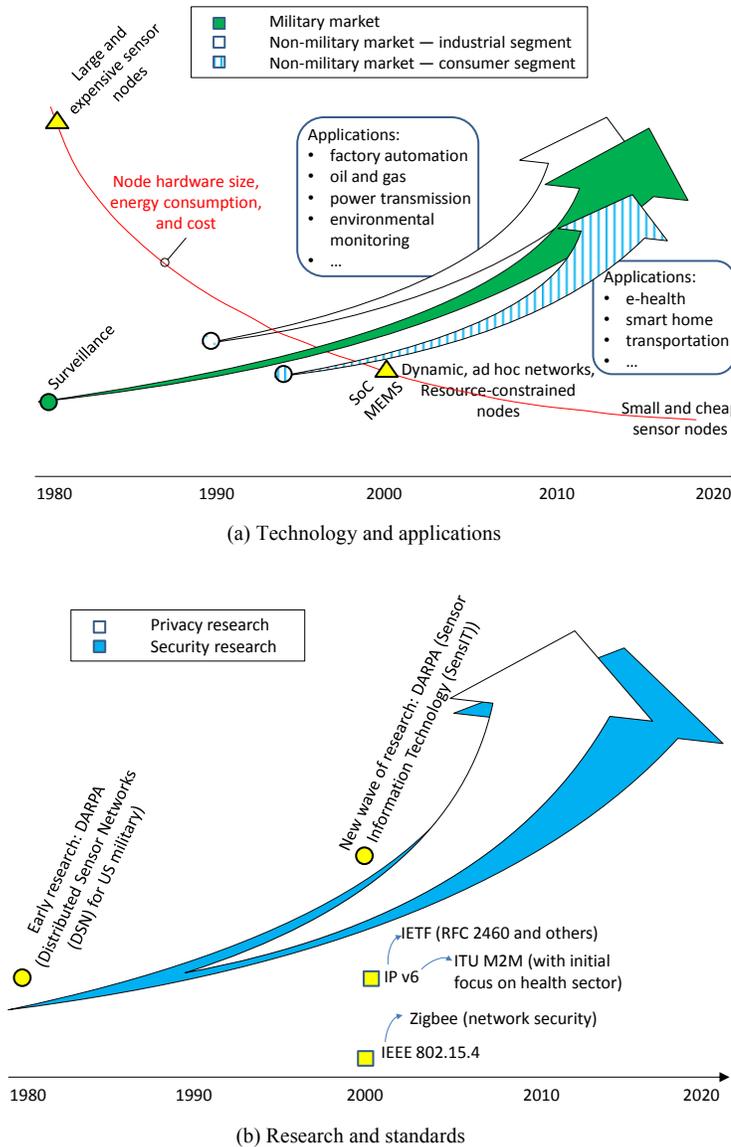

Figure 5. Timeline of significant developments and applications for wireless sensor networks

Singling out the monitoring of the elderly, as an illustrative example, wireless sensor networks enable real-time collection of physical and behavioral information about elderly people wherever they may be, at or away from home. This information can be used for analyzing and monitoring age-related diseases [33], for example. These types of applications for the elderly are gaining increasing relevance and importance, given the significant rise in life expectancy and the aging of the human population which the world is witnessing in recent decades [4].

It is predicted [4] that the worldwide population of people aged 65 and above will more than double to reach around 761 million by the year 2025. However, the capacity of the traditional healthcare system is not growing fast enough to cope with the predicted consequent rise of the pressure on healthcare provision [4]. This has given rise to the need for new ways of delivering the required healthcare while, at the same time, reducing costs [4].

As a significant development with regards to new ways of delivering healthcare, wireless sensor networks have already been deployed into the homes of some elderly people, to detect their daily activities and report unusual events [34]. For example, sensors have been attached to common household items (such as room, cupboard, or refrigerator doors; medicine boxes; key chains …) to track the movement and activities of the elderly [34]. The data collected from the sensors is uploaded, to be analyzed such as through comparison to previously-recorded daily routines and health-related histories and preferences. Healthcare professionals or family members can check the health status of the elderly by logging into a Web-hosted application (for example). They may examine the data collected by the sensors, possibly using automated data analysis functionality provided by some intelligent decision-support tool, to extract indicators of normality or abnormality. Abnormalities can be reported to a caregiver or to a loved one. It has been reported that such systems can improve the independence and the level of activity of the elderly, and they can help detect problems before they escalate into emergencies [34].

According to statistics released by the World Health Organization [35], the top of the list of leading causes of premature death worldwide includes heart diseases and strokes. The death rates could be reduced through appropriate healthcare, based on the collection of timely and reliable information about individual and public health, relating to such killer diseases. Wireless sensor networks can be used for collecting such information, anytime and anywhere.

Another possible application of wireless sensor networks relates to the millions of people in the world who suffer from chronic illnesses, like diabetes. The World Health Organization reported that diabetes is projected to be the seventh leading cause of death in 2030 [36]. Continuous monitoring of diabetic patients is very important, to ensure the proper dosing of medication and to reduce the risk of serious complications [37]. A similar requirement applies to many other illnesses. The deployment of wireless sensor networks in healthcare allows the continuous monitoring of patients and healthy individuals, thereby enabling preventive measures or rapid medical intervention, which can avoid illness or reduce disease-related complications and risks.

WSN-based healthcare systems can also enable the automated collection of better public-health data, such as reliable information on disease incidence rates and the number of births and deaths. The automation, timeliness and reliability of the collection of such data can enhance the formulation of treatment and well-being programs for individuals (hence, boost the personalization of healthcare), and the identification of national or transnational health priorities. Furthermore, the collected information can empower individuals to have more control over the health and social care which are delivered to them.

## 2.3 Applications of wireless body sensor networks

A wireless body sensor network is a group of small, lightweight, possibly intelligent, and low power wireless sensor nodes which can be placed on or inside a human body to continuously monitor the health of a human [38]. Wireless body sensor networks are, to a significant extent, based on general wireless sensor network technology. However, they impose challenging design considerations compared to traditional wireless sensor networks, particularly with regards to the scale of deployment, topology, data rates, power consumption, security level, tolerance of data loss or sensor loss, and the characteristics of the deployed sensors [38] [39] [40].

The nodes of wireless body sensor networks can be sensor nodes or actuator nodes [37]. The sensor nodes are responsible for measuring physiological readings. They are known as wearable sensors (carried on the human body) [41] or implanted sensors (placed inside the body) [37] [42]. Various types of biomedical sensors can be deployed in a wireless body sensor network, to wirelessly monitor physiological information relevant to the condition of a patient. Examples of wearable sensors are pulse oximeters (to measure oxygen saturation), electrocardiograms (ECG) (to monitor heart activity), thermometers (to record body temperature), blood pressure sensors, electromyogram (EMG) sensors (to monitor muscle activity), activity or motion detectors, and electroencephalogram (EEG) sensors (to monitor the activity of the brain) [43] [42]. Examples of implanted sensors or actuators are: glucose monitoring sensors, and implantable neural stimulators deployed to send signals to the human brain in diseases such as Parkinson's disease, for example [42].

Actuators perform a specific action based on the data collected by the sensors or according to user interactions. Such is the case of actuators equiped with an insulin pump and reservoir to administer insulin to a diabetic person [37]. Several types of body sensor hardware are available, including piezo-electric sensors which are used for pressure measurements, and optoelectronic sensors such as the infrared sensors used for body temperature estimation or heart rate or blood pressure monitoring [25].

The sensors of a wireless body sensor network can be integrated into a personal area network, by complementing the body sensor network with additional sensors carried by the person (portable sensors) or located in the private or public environment surrounding the data subject (surroundings-mounted sensors). These sensors may monitor either the person or the environment in which they are. Audio and video sensors (respectively, microphones and cameras) are a common example of sensors mounted in the surroundings.

### 2.3.1 General deployment architecture

Sensor networks are often configured into a multi-tier architecture, which uses multi-hop data transmission. Data are transmitted either to healthcare personnel or to a server within the information technology infrastructure used by the healthcare provider [25]. For example, [25] presents a three-tier architecture, which consists of: (i) a first tier made up of sensor nodes such as wireless body sensors, possibly grouped into clusters; (ii) a second tier of network connector nodes, which include cluster heads; (iii) a third tier consisting of the base station node.

The sensor nodes acquire the data of interest. To ease network deployment, they typically communicate wirelessly within a defined region. They can be line-powered or battery-operated, and they are usually deployed in a cluster. Sensor nodes may relay data to transport nodes; usually, these are lightweight nodes with a low monetary cost. Often, they are wireless and battery powered, and are typically deployed in a cluster, whereby a special transport node in the cluster communicate with the base station. This special node is known as cluster head. The base station often has significant computational and communication capabilities, and is conventionally assumed to be a central trusted authority.

Several prototypes have been developed for the deployment of wireless body sensor networks. They mostly follow a multi-layer or multi-tier architecture [44]. Figure 6 is a general outline of the basic deployment architecture for a wireless body sensor network in a healthcare system.

Tier 1 would typically consist of the Body Area Network (BAN) and the Personal Area Network (PAN) [15]. The BAN is made up of the wearable and/or implanted sensors and actuators, which are used for monitoring physiological signs or for drug administration. The PAN consists of other sensors which can be deployed around the patient, such as sensors mounted in the surroundings of the patient. In Tier 1, each sensor is responsible for sensing, sampling and processing the relevant physiological signal [45].

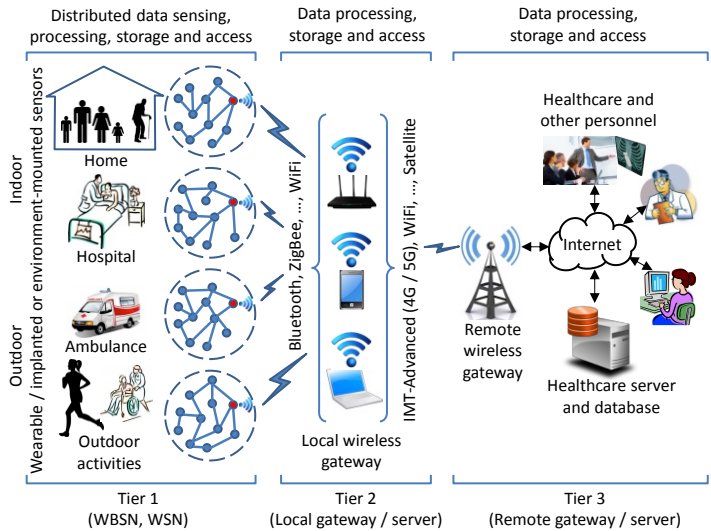

Figure 6. Simplified system architecture for healthcare applications based on wireless sensor networks (adapted from [15] [19] [25] [7])

Tier 2 is the personal server stage which can be made up of a so called 'personal digital assistant', a home computer, or a cellular telephone; it is responsible for interfacing with the network(s) in Tier 1 and with the medical server(s) in Tier 3 [45] [44]. The interfacing with the personal server includes the network configuration (sensor node registration, configuration and security setting) and management features (retrieval and processing of data, scheduling, channel sharing and data fusion) [45].

Tier 3 is the medical server tier, which may include medical emergency servers and links to caregiver devices [44].

In [46], the wireless sensor network is similarly configured as a hierarchical topology. The network consists of sensor nodes, transport nodes, and a base station; they are organized into four layers. In the lowest layer, camera nodes are grouped into clusters, each with cameras which capture correlated visual readings. Each camera node feeds visual data to transport nodes which are in the second layer. Transport nodes form multi-hop transmission paths through which data are relayed towards special transport nodes (the cluster heads); there is one independent path and cluster head for each camera. These cluster heads make up the third layer. They communicate with the base station (which is effectively the fourth layer of the topology).

### 2.3.2    Illustrative application scenarios

Wireless body sensor networks have found applications in numerous aspects of life. As depicted in Figure 7, applications of wireless body sensor networks can be categorized into healthcare, assisted living, and others (gaming and entertainment, military applications and emergency services)

[47] [39].

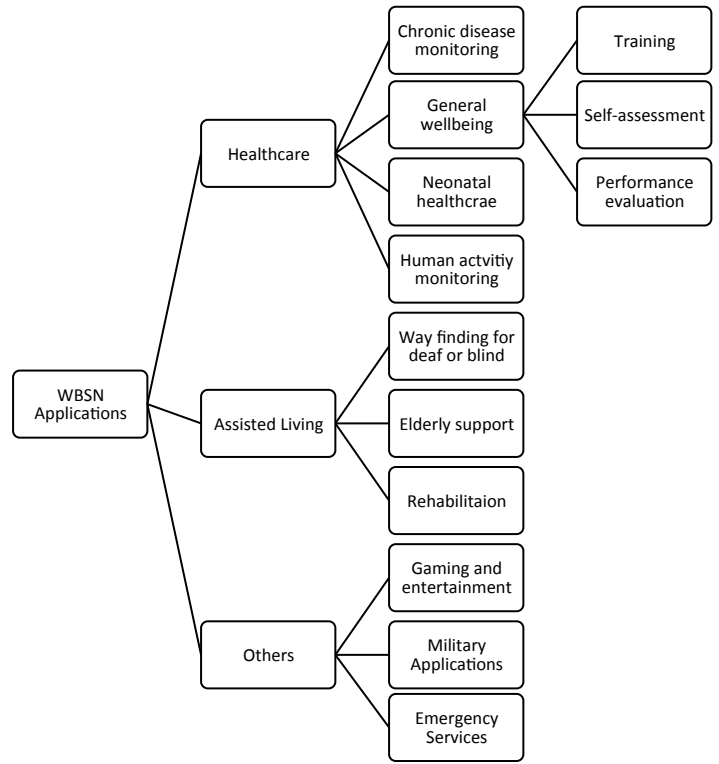

Figure 7. Applications of wireless body sensor networks (adapted from [47])

Healthcare applications can be further categorized into chronic disease monitoring, general wellbeing, neonatal healthcare and human activity monitoring. In chronic disease monitoring, a wireless body sensor network can be used in the fight against cardiovascular disease, by monitoring patients remotely in real time, to enable them to have a healthy lifestyle and facilitate the early prediction of emergencies [47]. According to [47], the use of wireless body sensor networks to achieve general wellbeing can be valuable in the development of coaching systems, self-assessment, continuous monitoring and in the performance evaluation of a normal human being. They can also be used in the training of athletes, dancers and other performers. In neonatal healthcare, wireless body sensor networks can be used for the continuous and non-intrusive monitoring of newly born babies. Wireless body sensor networks can also be used for older children or adults, such as for detecting infectious diseases or chronic health issues, and for monitoring healthy living habits [47].

In assisted-living, wireless body sensor networks can be deployed within way-finding tools for the blind or deaf, in real-time activity monitoring for elderly people and for the physical rehabilitation of injured patients. In way-finding, wireless body sensor networks can help blind people to move around in familiar or new environments. Furthermore, visually-impaired people could have an artificial retina (made up of micro sensors) implanted in the eye, to generate neurological signals based on a camera mounted on eye glasses [37], and these micro-sensors can be connected to a wireless body sensor network.

In the context of rehabilitation, continuous remote monitoring of patients (such as those who suffered a stroke or underwent surgery, or suffer from motor dysfunction) can be used to non-intrusively assist in the recovery process for these patients. Wireless body sensor networks can also be used for the training of patients suffering from motor impairment and for physical exercise instruction targeted at the elderly [47], for example. Wireless sensors could be attached to the legs or the nerves, and actuators could be used to stimulate the nerves, so as to assist in the required motion of the limb or body [37]. Wireless body sensor networks can be used to support the independence of elderly people living on their own, by detecting their activities such as walking, sleeping, and even falling or other accidents [47].

Another possible application for wireless body sensor networks is the remote monitoring of soldiers on the battlefield, to measure their fatigue level, posture and vital signs. In the context of emergency services and extreme situations, wireless body sensor networks can be used to detect toxicity in the air and warn fire fighters, for example [37].

Other application scenarios which have been suggested [47] are: identifying frostbites for people subjected to very cold weather conditions (such as swimmers, soldiers and outdoor workers); assistance for visually-impaired people engaged in sports such as swimming; and miniature detection-and-alert systems for monitoring the fluctuation of blood sugar levels for diabetic patients.

## 2.4    Privacy issues

### 2.4.1    The privacy protection imperative

As illustrated in the foregoing sections, wireless sensor networks offer a wide range of actual and potential applications in healthcare. However, the privacy of the subject whose data is collected by sensors and the privacy of all other stakeholders in the healthcare system are of crucial importance in the applications discussed above. Privacy must be preserved adequately [48]. Hence, the acceptance of wireless sensor networks technology by stakeholders in healthcare applications will be influenced by the effectiveness of the techniques deployed to safeguard the personal and possibly intimate information which can be captured and transmitted over the air, within and beyond these networks.

It is highly important to maintain the privacy of the information flowing through a wireless sensor network for healthcare. Without privacy preservation, these systems cannot be accepted by people, regulatory bodies, or governments. Privacy preservation is thus an important consideration in the development and deployment of wireless sensor networks for healthcare applications. Furthermore, technological developments are not sufficient on their own; they have to mesh in smoothly with the legal framework within which the technology will be used, given that the law will have a significant bearing on the operational requirements that privacy-preserving wireless sensor networks technology will have to meet, in order to be ready for deployment in the real-world.

Pfitzmann and Hansen [17] proposed a terminology, in relation to privacy protection goals which are required for achieving privacy by data minimization. Here, data minimization refers to the minimization of the possibility of collecting personal data about others, the minimization of actual collection of personal data, and the minimization of how long collected personal data is stored, such that "No subject should get to know any (potentially personal) data – except this is absolutely necessary" [17]. The five key privacy goals, defined in [17] and commonly encountered in the literature in the field privacy by data minimization, are: anonymity, pseudonymity, unlinkability, undetectability and unobservability. Their definitions are included in the appendix.

A system (such as a communication network, for example) which protects the privacy of its users should meet the applicable privacy protection goals, by deploying appropriate Privacy Enhancing Technologies (PETs). In effect, Pfitzmann and Hansen [17] defined a terminology for special privacy protection goals which extends the classical goals of secure systems. According to the security model known as the "CIA" triad, these classical security goals are: *confidentiality* (information is not accessible or disclosed to unauthorized parties or processes), *integrity* (information is kept accurate and complete) and *availability* (information is accessible by authorized users, when needed). Other privacy protection goals have been advocated in the literature [16].

### 2.4.2    The multifaceted challenge of privacy protection in wireless sensor networks for healthcare

Privacy protection is recognized as mandatory for healthcare applications, it is however a challenging task because of the complexity which arises from the many factors which affect the degree of control that a subject can exercise over their personal health data (see the definition of privacy which is given in the appendix), as illustrated in Figure 8. There are many possible combinations of multiple factors which make it difficult to protect against infringement of the rights of the data subject to control the acquisition, storage and use of their personal health data. Indeed, with regards to data collection, there are many possible *contexts* within which the data can be collected, coupled to several possible *locations* of sensors, wide variety of *types of personal data* and *formats* of sensor data. Furthermore, the data access or disclosure dimension introduces a wide range of potential *data accessors and users*.

Concerning the spatio-temporal context of the data collection, data can be collected within personal spaces (where the subject would normally have significant control over the data collection) or within public spaces (often characterized by a minimal degree of control that the subject would exert over the data collection). In addition, data can be collected sporadically or continuously, potentially over a very long period of time, possibly resulting in a high amount of data which could be stored in disparate locations by different entities, the existence of which the subject might not even be aware of. The types of personal data can include physiological and mental health data, and other personally identifiable data relating to lifestyle habits, physical activities and social interactions, for example.

Another axis contributing to the complexity is that the collection of data can be over a long time span, which could be years in the case of chronic conditions. The volume of personal data may thus be immense, and the corresponding health records may be stored as Electronic Health Record (EHR), or as Personal Health Record (PHR). These records may reside with different healthcare providers, possibly across national boundaries, and the collected personal data can be disclosed or accessed or used for a wide range of purposes (including primary and secondary ones), by the data subject or by others such as registered service providers, social or family connections, employers, or research or law enforcement organizations. The data subject may have little control over these records, particularly with regards to secondary use. Moreover, different situations may require different privacy settings or levels of compliance with the choices of the data subject. For example, a medical emergency may require the privacy choices made by the data subject to be ignored.

As discussed earlier, the architecture of the communication system could consist of multiple tiers, with devices possibly located in different places under the control (if any) of different people, authorities or organizations. Personally identifiable health information flows from its acquisition by the sensors, possibly followed by local processing or temporary storage at the sensor node, and by wireless multi-hop transmission to the sink node. The information is possibly subjected to further processing and temporary storage at the sink node, and to eventual transmission from the base station (see Figure 6 and Figure 9). There are several potential vulnerabilities along the information flow path.

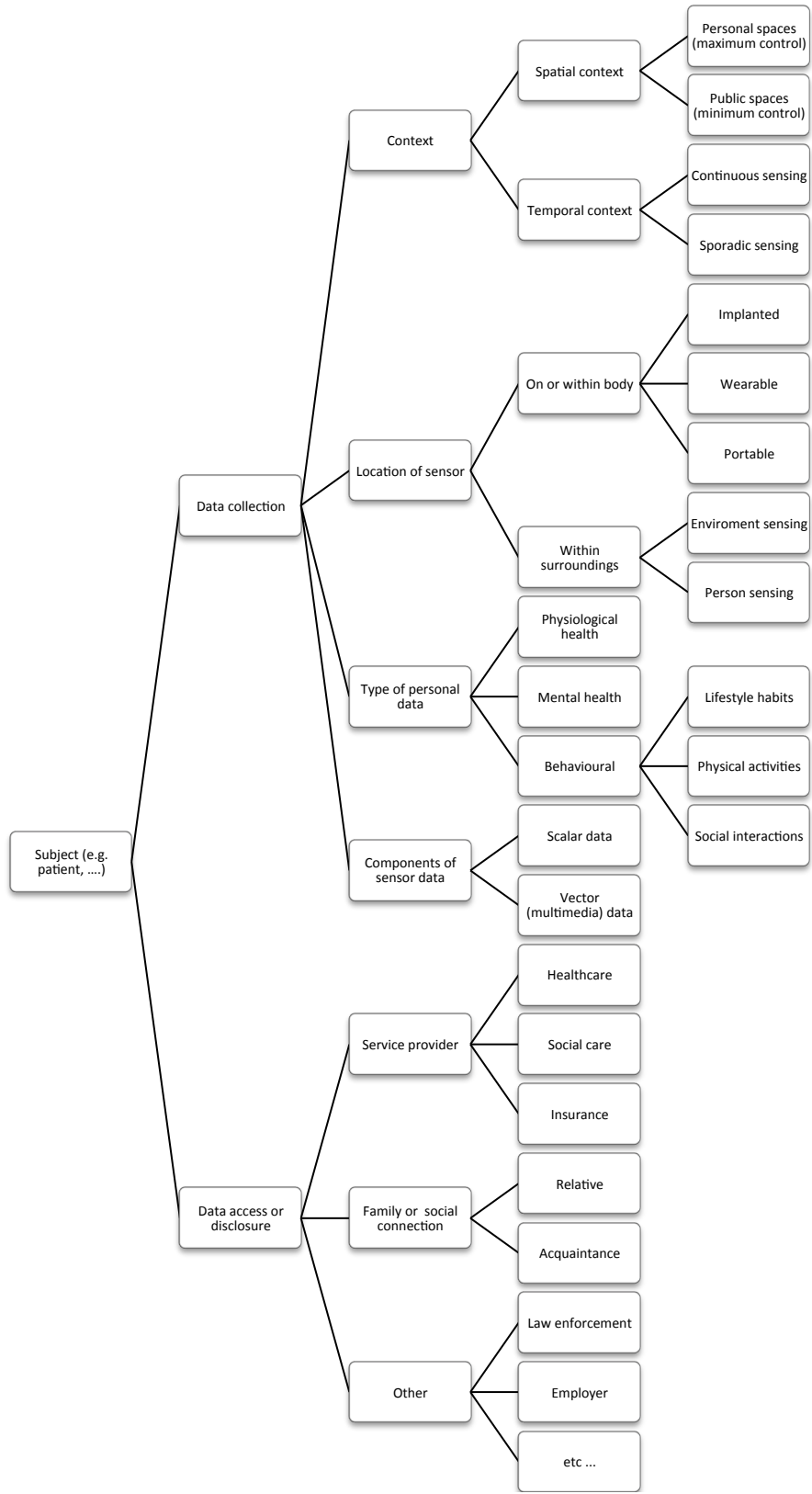

Figure 8. Factors associated with the multifaceted challenge of privacy protection in healthcare. The degree of control that a subject can exercise over their personal health data is affected by the interplay between many factors which relate to data collection, access or disclosure.

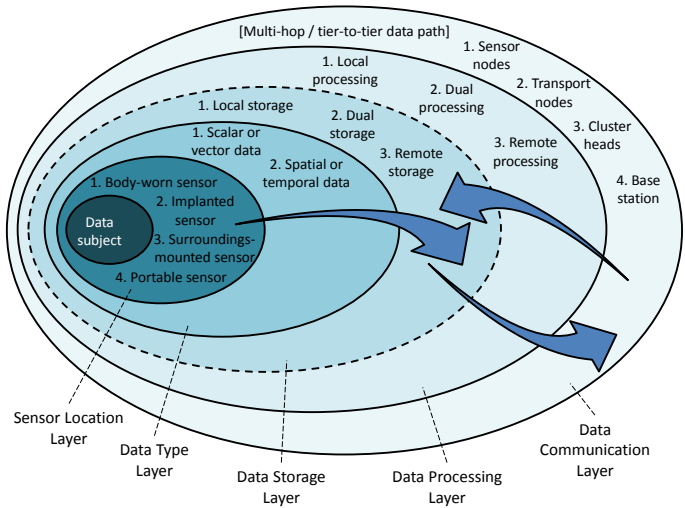

Figure 9. Layered representation of the flow of private data in wireless sensor networks for healthcare. The broken line between the storage and processing layers indicates that the sequence of the two layers along the flow of data could be interchanged or even loop between the two layers.

For example, sensors or other nodes may be stolen or get lost (devices are possibly tiny; hence there is a high risk of device loss). Theft or loss may lead to unwanted data access or tampering by unauthorized parties. Furthermore, as discussed earlier, the sensors can be implanted, wearable, portable, or mounted in the surroundings; some sensors (such as cameras) usher in a significant potential for intrusion (by outsiders, malicious or otherwise) into the private life of the data subject. In addition, the sensor and other nodes typically operate and communicate wirelessly, in environments which may be difficult to control, within clinical or other spaces. Data transmitted wirelessly may be intercepted or tampered with. Furthermore, the wireless sensor network may consist of an eclectic heterogeneous collection of sensors or communication devices; with devices swapped in or out of the network dynamically. For example, patients moving from a clinical setting to a non-clinical setting may change sensors, or sensors may be shared by successive patients at a clinic. In addition, the variety of personally identifiable data may enable an attacker to combine multiple types of information to re-identify the data-subject from anonymized data.

Another dimension to the challenge is that there is a variety of privacy policies and laws, across regional or national boundaries; hence there may be a significant variation in the requirements to be met by privacy enhancing technologies which support the protection or enforcement of privacy rights. The laws may also impose restrictions on flows of data across jurisdictional boundaries.

The primary utility of wireless sensor networks in healthcare is to collect and make the relevant information accessible to service providers or other stakeholders. Privacy and security techniques are thus an additional pressure on the often constrained hardware resources and on the corresponding energy constraints. Achieving the privacy and security goals of wireless sensor networks for healthcare, while using constrained hardware resources, often gives rise to a tension at run-time between the achievable level of privacy or security and the hardware-related costs — processing, storage and communication, and consequently energy costs (see Figure 10). The level of attainment of each individual privacy or security goal is often adjusted, as demanded by the requirements and constraints of the situation at hand, and as determined by the nature of the privacy or security service(s) deployed for the given situation. The tension, between costs and the level of privacy or security, often imposes trade-offs such as the use of lightweight privacy and security algorithms, possibly affording a lower level of privacy protection than would have been the case in the absence of resource constraints. The problem of constrained hardware resources is particularly significant for wireless body sensor networks.

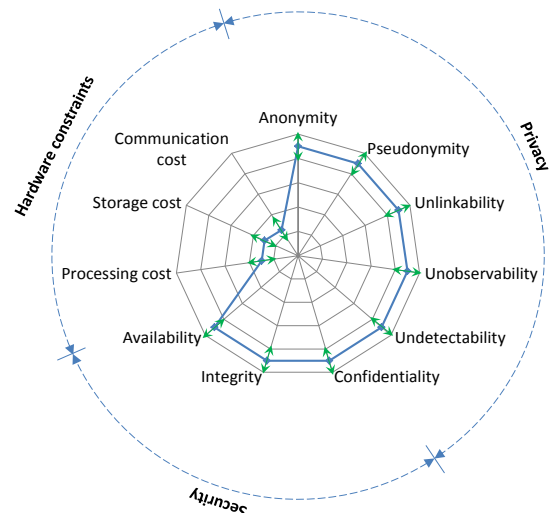

Figure 10. Tension between hardware constraints and privacy and security goals for wireless sensor networks in healthcare. Ideally, the level of attainment of privacy and its related security goals should be maximized and run-time hardware resource costs minimized. However, in practice, each privacy or security goal is often adjusted, as demanded by the situation at hand and as determined by the nature of the privacy or security service(s)

deployed for the given situation. Trade-offs are often required between the level of privacy protection and costs.

By virtue of its nature, healthcare data often requires collection, access and use by others, for medical or other purposes. Hence, sophisticated technical tools and legal instruments are needed in order to preserve the privacy of the data subject, under the operational constraints of wireless sensor networks. The deployment of the privacy and security services in a WSN-based healthcare system should protect all stages of the system. These stages include data capturing, communication, processing and storage; and the data could be at different levels of abstraction, from raw data to high-level semantic information (see Figure 9). In general, privacy services should be applied to the data captured by sensors, data transmitted to the sink, data processed in the sink, data transmitted to the gateway, and to data being processed and stored on remote servers (this last stage is out of the scope of this paper).

## 3   PRIVACY SERVICES IN WIRELESS SENSOR NETWORKS

This section discusses privacy services required for the deployment of wireless sensor networks in healthcare. It begins by offering a definition of privacy service and privacy mechanism. Then, it discusses privacy gaps in early work on healthcare systems which are based on wireless sensor networks. Thereafter, it reviews developments beyond the early work, with a focus on services and their underpinning mechanisms to achieve privacy goals explicitly or to protect location privacy, which is an important requirement in healthcare applications.

### 3.1   Definition of privacy service and privacy mechanism

The survey papers discussed in Section 1.4 adopt different terminologies in relation privacy. For example, some use the terms 'privacy requirements' [7] or 'privacy goals' [8], and others refer to them as 'privacy problems' [26].

This section adopts definitions of services and mechanisms, which are borrowed from information security, and it applies them to privacy protection. According to [49], a security *mechanism* is defined as "A process (or a device incorporating such a process) that is designed to detect, prevent, or recover from a security attack", and a security *service* is explained as "A processing or communication service that enhances the security of the data processing systems and the information transfers of an organization. The services are intended to counter security attacks, and they make use of one or more security mechanisms to provide the service". In [49], security services are subdivided into authentication, access control, data confidentiality, data integrity, and non-repudiation services.

Extending to privacy the concept of "service" which is explained above, this paper defines a privacy service as 'a software component which provides privacy-enhancing functionality to other components of a system'. In the context of this paper, the said system is a healthcare system. A service may be invoked on its own or combined with other services to provide privacy protection. The service harnesses privacy-enhancing mechanisms to fulfil the relevant privacy requirements or goals for the system. This paper thus defines a privacy mechanism as 'a processing technique (or a device which incorporates such a technique) which is tasked with handling privacy attacks'. Specifically, a privacy mechanism 'can detect, prevent, or recover from a privacy attack'; in a similar vein to the definition of a security mechanism, which is given above.

Thus, privacy services can be defined at different levels of granularity along the taxonomy presented by [26]. For example, services can be deployed as data privacy services (targeting data aggregation and query privacy), and contextual privacy services (directed at location and temporal privacy). Furthermore, privacy services can be deployed at the level of privacy goals, namely: anonymity, pseudonymity, unlinkability, undetectability, and unobservability services. Thus, Section 3.3 is structured into reviews of mechanisms, which can be used in lower-level privacy services for privacy goals, or in higher-level services for data-oriented or context-oriented privacy. Privacy mechanisms to tackle attacks are discussed in the context of privacy services, and they are summarized in TABLE II.  and TABLE III.

### 3.2   Privacy gaps in early work on healthcare systems based on wireless sensor networks

This section discusses the early generation of healthcare projects, including highly cited projects in the literature [4], these are: CodeBlue [50], Mobicare [51] and SATIRE [52]. With regards to privacy and security considerations, it should be pointed out that only a few among the early generation of healthcare systems (such as ALARM-NET [53] and MeDiSN [54]) embedded some security services, and they did not consider privacy in its own right, as they viewed privacy preservation as a by-product of the deployment of security services. This section presents a quick review of early healthcare projects, along with a table summarizing the privacy and security services suggested or implemented for these systems.

The CodeBlue [50]  architecture used the Elliptic Curve Cryptography technique on its MICA2 motes to ensure security of data transmission. However, the encryption key required thirty-two seconds to be generated, which was considered unsatisfactory [55]. An extensive security threat analysis was conducted on the CodeBlue architecture, and it was shown to be vulnerable to security attacks such as denial-of-service attacks, snooping attacks, modification attacks, routing-loop attacks, grey-hole attacks, Sybil attacks and masquerading attacks [5][56]. Security attacks on the CodeBlue architecture may have serious impact on privacy. For example, in case of a snooping attack, an adversary might acquire private information about patients by observing the operation of the relevant components of the physical system. A Sybil attack may result in incorrect decisions based on false health information sent to the sink node. A masquerade attack might have an impact on privacy due to gaining access to the whole system using stolen user identifiers and passwords. Consequently, the CodeBlue system suffers from serious privacy vulnerabilities which can prevent it from becoming a WSN-based healthcare system in the real world.

The developers of the Mobicare [51] architecture suggested the use of the Wireless Transport Layer Security (WTLS) protocol to provide patient privacy, data integrity and authentication. However, the WTLS was not actually implemented or tested on the Mobicare architecture [5]. This makes the Mobicare architecture vulnerable to all possible privacy attacks

The SATIRE [52] project did not implement any of the suggested security and privacy services and considered them as future work [5]. The developers of SATIRE suggested the use of an access matrix to preserve the privacy of data in their system. The main idea of this access matrix is to define who can access what. However, this basic security scheme would not be enough to stop privacy attacks. Restricting access to the data of the patient only to authorized personnel or family would not stop adversaries from invading the privacy of the patient. As a potential negative impact of lack of anonymity or pseudonymity services, curious employees or families may access, publish or even sell critical medical information to employers or insurance companies, which might have serious implications for the patient. In addition, as a result of lack of a location privacy service, continuous monitoring by adversaries might expose the geographical location of the patient; such a breach might not be acceptable by many patients.

The developers of the MeDiSN [54] architecture highlighted the need for encryption, to ensure the confidentiality and authenticity of the delivered data. However, they did not reveal any details about the security authentication or cryptosystems used in MeDiSN [5].

The Alarm-Net [53] architecture used a secure remote password protocol for user authentication for IP-network security. In addition, the sensors (such as MicaZ and Telos) used in this healthcare system had built–in cryptosystems. Although Alarm-Net offers both authentication and encryption operations, it suffers from major drawbacks. The cryptosystems used are highly platform-dependent. In addition, they do not offer decryption options, which denies intermediate nodes access to the data during communication [5].

Although, the use of both authentication and encryption grants lawful personnel access to the system, and prevents eavesdropping on the traffic, it does not grant privacy protection in relation to location privacy, for example.

TABLE I.     SUMMARY OF THE PRIVACY AND SECURITY SERVICES SUGGESTED OR IMPLEMENTED IN THE EARLY WIRELESS SENSOR NETWORKS FOR HEALTHCARE SYSTEMS

| Project Name | Services suggested in the relevant papers | | Implemented services reported in the papers | |
|---|---|---|---|---|
| | Security Services | Privacy Services | Security Services | Privacy Services |
| CodeBlue | Authentication, cryptography | None | None | None |
| Mobicare | Shared Keys, authentication | None | None | None |
| SATIRE | Authentication | Access Matrix | None | None |
| ALARM-NET | Authentication, cryptography and key management | Dynamic Privacy | Authentication, encryption | Dynamic authorization to access patient vital information |
| MeDiSN | Authentication and cryptography | None | Not described | None |

TABLE I. summarizes the privacy and security services and mechanisms, which were implemented or proposed for the architectures and systems developed in early work on healthcare systems based on wireless sensor networks.

Based on TABLE I. , it is evident that the early incarnations of WSN-based healthcare solutions did not focus on the importance of implementing privacy services in their systems. None of these healthcare projects included a thorough analysis of privacy attacks, or studied adequately the required privacy services for defending against attacks.

Since all the projects mentioned above are not very recent, other recent projects were also studied, to check whether privacy services were among the system considerations. It was found that the publications on WSN-based healthcare systems such as KNOWME [57], and those presented in [58] and [59], did not mention security or privacy services.

### 3.3 Developments beyond the early work

#### 3.3.1 Services and mechanisms to achieve privacy goals

Many privacy techniques developed for wireless sensor networks in general can be adapted to the requirements and constraints of healthcare systems. However, particular care is required, to deal with the limited resources associated with some applications, such as those applications which require wireless body sensor networks.

Given the pivotal role of privacy goals, for achieving privacy by data minimization, this section classifies privacy enhancing services in relation to the five key privacy protection goals which were defined by Pfitzmann and Hansen [17], namely: anonymity, pseudonomity, unlinkability, undetectability and unobservability.

##### 3.3.1.1     Anonymity

Preserving anonymity in wireless sensor networks requires multifaceted solutions, which address data anonymity, device anonymity and communication anonymity [60]. Hence, the anonymity service has been addressed in many papers, along many dimensions of anonymity such as: base-station anonymity; source anonymity; user anonymity; query anonymity; data collection anonymity; and communication anonymity.

**Base station anonymity:** This denotes the hiding of the identity, role and location of the base station from external adversaries [61]. Attacks on the base station can have a debilitating effect on the network because the base station is a critical part of the network, being the sink and convergence point of all traffic [62]. A variety of approaches have been developed to try to protect the anonymity of the base station against malicious attacks. Most of these approaches rely on creating a perception that the base station is a typical sensor node. Two approaches for base-station anonymity are suggested in [61].

The first approach is based on making the base station transmit messages to random nodes in its neighborhood. The neighborhood nodes retransmit these messages to nodes away from the base station, thus creating the illusion that the base station is just another sensor node in the network. The second approach is a line of defense against long-time traffic analysis which may be used by adversaries to eventually identify the base station. In this approach, the base station may be relocated when motion is possible. However, the relocation of the base station must be carefully analyzed to calculate the threat level and implications of the relocation [61].

Another approach, suggested by [62], for increasing the anonymity of the base station, is to increase the transmission power of the nodes of the network to achieve longer transmission ranges, which increases the correlation between neighboring nodes and makes traffic analysis very complex. Although this approach can avoid changes in routing protocols and in traffic patterns to deceive adversaries, the increased transmission power may have a serious effect on the network life-time and on interference between signals transmitted by different nodes [62].

Recently, [63] suggested the use of beamforming, to boost the anonymity of the base station while minimizing the communication energy overhead. In [63], distributed beamforming by nodes with single antennas cooperate to form a virtual multi-antenna system, to improve the communication range, data rate, energy efficiency, and security of the physical layer, and to decrease signal interference. The distributed beamforming technique is deployed as three components: a cross-layer relay selection algorithm, to determine which nodes will participate in the beamforming; a time synchronization algorithm, to construct a common time reference; and a carrier synchronization algorithm, to create a common frequency reference.

All approaches described above are based on *hiding the identity* of the base station and attempting to make it appear like a sensor node. Other techniques aim to *disguise the location* of the base station. These are countermeasures against an adversary who uses traffic analysis in the form of packet tracing to locate the base station [64]. Traffic analysis to locate the base station is based on the idea that the traffic volume near the base station tends to be bigger than that away from the base station, which makes the location of the base station deducible based on the volume of traffic [64]. For example, packet tracing is deployed by an adversary to learn the hop-by-hop transmission links of nodes towards the base station [64]. It is claimed in [64] that packet tracing is more efficient than other forms of traffic analysis to deduce the location of a base station. Countermeasures against traffic analysis are reported in [65] [66] [64] [67] [68].

The work reported in [65] proposed four techniques based on randomized traffic volumes as a defense against traffic analysis, to protect the location of the base station. The techniques are: multi-parent routing; random walk; random fake paths; and fractal propagation. Multi-parent routing relies on the random selection of one of the parent nodes (connected to the node) to forward the data to the base station. This makes it hard for an adversary to detect a pattern to lead to the base station. In the random walk technique, the node forwards packets to its parent nodes, based on a random forwarding algorithm, thus distributing the traffic of the packets and decreasing the effectiveness of rate attacks. The random fake paths technique introduces fake paths on the way from the node to the base station, to reduce the effectiveness of time correlation attacks. Finally, the fractal propagation technique is based on the creation and propagation of fake messages into the network, to create areas of high activity and randomness in the communication pattern, to defend against rate monitoring attacks. In the fractal propagation technique, a node neighboring another node which is forwarding a packet to the base station, generates a fake packet according to some probability of creating packets, and forwards it to one of its neighbor nodes. The transmission paths of fake packets form a tree-like pattern in the network.

In addition to the randomness-based approach presented in [65], other countermeasures aiming at hiding the location of the base station are suggested in [66]. The suggested countermeasures include the use of: hidden packet destination address; de-correlating packet sending time; and controlling packet sending rates. Hiding the packet destination address is done through the encryption of the packet destination address, packet type and content, to hide the final destination of the packet (which is the base station). De-correlating sending time is achieved by introducing a random delay between the sending and the receiving of the packets between the parent and child nodes, to try to prevent the adversary from learning the hierarchy tree of the network. Controlling packet sending rates is achieved by creating a uniform traffic volume in the entire network. However, all these countermeasures may limit the data sending pattern of the network, hence they may not be suitable for situations when urgent data needs to be sent to the base station quickly. They also increase energy loss (due to the use of dummy packets) and increase the overall delay of the network (due to the introduction of random delays). The authors claim that the methods introduced in [65] outperform those in [66].

Similarly, other techniques suggested in [67] and [68] are proposed as countermeasures against packet tracing, they use a location privacy routing protocol combined with injection of fake messages. The main idea of [67] is to randomize the routing paths towards the base station and inject fake messages into the network, to uniformly distribute the incoming and outgoing traffic at a sensor node. Although their scheme aims at hiding the location of the base station, the trade-offs between location privacy and energy consumption and network delays should be scrutinized. The injection of fake messages, sent towards the base station using biased random walk combined with a routing table perturbation scheme which modifies the routing tables of the nodes, was claimed to be robust against local adversaries [68].

Another method to hide the location of the base station is proposed in [64], using two methodologies to hide the location of the base station during topology discovery and data transmission. An anonymous topology discovery scheme is used to conceal the location of the base station. In this scheme, a common sensor node, pseudo-base station, is randomly chosen by the base station to pretend to be a base station and initiate a topology discovery. During the data transmission phase, the base station location is concealed using a fake message injection scheme. This scheme is based on the idea that a sensor node will transmit fake messages to the neighboring nodes at the same time that it is forwarding real packets, which would cause the adversary to waste time studying fake paths. This scheme is combined with a simple version of the random walk algorithm to hide the location of the base station.

**User anonymity:** Allowing users to directly access sensor nodes to obtain real-time data requires considerable privacy and security measures to protect critical data [69]. Accordingly, user authentication is a crucial mechanism to grant access for rightful users [69]. A variety of techniques are available to anonymously authenticate users who require access to the sensor node data. For example, authentication based on a smart-card and a password (two-factor authentication) is deployed in user authentication mechanisms due to its simplicity, portability, and security [69]. Several research investigations have been reported in the literature, to achieve user anonymity through two-factor authentication in wireless sensor networks. Some of this research was intended for use in general wireless sensor networks and others were intended for WSN-based healthcare system. A survey of the two-factor authentication schemes for wireless sensor networks is presented in [69], with an assessment of previous attempts to design user anonymous two-factor authentication schemes. In addition, they present their own scheme which is a combination of the schemes of Fan et al. [70] and Xue et al. [71].

A user anonymity technique, for the Smart CArd based user authentication scheme for WSN (SCA-WSN), is presented in [72]. In this scheme, a user holding a special smart card issued by the gateway can gain access to the sensor data after authentication from the gateway. The basic idea of the scheme is to deploy elliptic curve cryptography which is used only when there is a user-gateway authentication to anonymously authenticate users to access the sensor node data.

Although anonymity has been investigated, or just proposed, for WSN-based healthcare systems, details of the underpinning techniques are often missing, such as in [73] [74]. To the best of our knowledge, the main focus of investigations conducted on anonymity for WSN-based healthcare systems has thus far been on user authentication anonymity. For example, [75] proposes a biometric authentication-based protocol for ensuring the privacy of the patient. The paper claims that the protocol enhances the anonymity level, and it is computationally more efficient, compared to others. Other papers on user anonymity for WSN-based healthcare systems are [76] [77].

**Query anonymity:** Some wireless sensor networks are designed to be owned and deployed by the same organization, while others are designed and deployed by more than one organization and may even extend over more than one country [78]. Clients who issue queries to such wireless sensor networks may require the anonymization of their interests and query patterns, which brings the need for private and efficient queries [78]. In addition, query anonymity can be of critical importance in situations where an adversary can deduce — based on the increasing number of queries to a specific location — where a patients dwells or whether a person might have health problems [26]. Several techniques have been developed to provide query anonymity. For example, two trust models are investigated in [78]. The first model consists of multiple, mutually distrusting servers which govern the sensor network. In the second model, all queries are performed through a single server, and they are sent unencrypted to the server.

For the first model, a protocol is proposed to ensure full client query privacy in a network with honest but curious non cooperative servers [78]. The protocol, called SPYC, divides the interaction of the client with the sensor network into two tasks: private naming of each sensor and private accessing of the readings of the sensor nodes. The first task, which is performed once for each client, builds a virtual naming space as a mapping of actual identifiers of the region where each sensor is located. The second task performs region-based source routing; the task is executed once for each client query, it privately accesses sensors of interest through the sensor network by using routing which is based on virtual region names. Furthermore, the client uses cryptographic techniques to hide from the servers both the virtual sensor names and the actual sensors which have been queried. In essence, the query is obfuscated using virtual region names and cryptographic techniques.

The SPYC protocol is made up of four procedures: initialization, key-space generation, query routing, and results routing. The main idea of the SPYC protocol is that the clients use the key-space generator to produce fresh keys, shared with the sensor networks and unknown to the servers. The clients use these fresh keys during the routing of the query and its result, to create packets which will be routed through the network and will be interpreted only by the designated nodes.

The privacy protection algorithms proposed in [78], for the second model, use another form of obfuscation, whereby the client query is expanded to include regions (sensors, by implications) beyond the one(s) queried by the client. This expansion is performed using a transform which produces a set of regions, when given a region of interest (as specified in a client query). The server will then query this expanded set of regions. Two privacy metrics were used to quantify the privacy level (spatial privacy level and temporal privacy level) respectively relating to the ability of the transform to hide the spatial and temporal patterns of the original query.

In [79], the Distributed Privacy Preserving Access Control ($DP^2AC$) is proposed. It ensures that access to sensor data is anonymous. The main idea of this scheme is that a client willing to access the data of a sensor network must first buy tokens from the network owner. Once the token is validated, the client is able to access the data which is required. To ensure the anonymity of the client, the token generation involves blind signatures, which can be validated by the sensor nodes and at the same time cannot identify the token holder. This scheme ensures both the protection of the privacy of clients and prevents the unauthorized access of the sensor nodes. Their proposed scheme has also a Distributed Token Reuse Detection (DTRD) scheme to ensure that sold tokens are not reused by malicious users.

Yi et al. [80] propose a method to enable secure distribution of patient data in multiple data servers, and protection of the privacy of patients by using the Paillier and ElGamal cryptosystems when statistical analyses are performed on patient data. This method aims to protect the patient data against an attack by an insider such as an administrator of the database containing sensitive patient data.

It is important to mention that the trade-off between the communication costs and the anonymity of queries is a critical issue that was discussed in [81] [78]. It is desired that the private query mechanism should have minimal overhead above a non-private query. For example, [78] offers algorithm parameters which can be varied to achieve trade-offs between privacy and efficiency levels.

**Source anonymity:** Source anonymity is a critical and challenging task where the source node reporting a certain event must be guarded from adversaries to protect it from being identified [82]. Several solutions have been developed to protect the anonymity of the source. The solutions are based on the utilization of fake messages and are aimed at global adversaries monitoring the whole network traffic, as shown below.

A scheme called FitProRate is proposed in [82], to achieve source anonymity. The main idea of the scheme is to use dummy messages such that the sensor nodes maintain a constant traffic pattern in the network. When a real event is detected, the node waits to send the real packet during the same time slots which it uses to send the dummy ones. This makes it hard for an attacker to detect the real source of the event. However, this mechanism introduces latency into the network. In order to try to decrease the overall latency, the authors suggest the adoption of an exponential probability distribution to determine the time interval to use for sending messages within the network. A similar approach is presented in [83][84]. In addition to the basic source anonymity approach, further studies are presented by the authors to analyze the intervals for transmission times of the fake and real packets, to try to defeat traffic analysis by an adversary and decrease the overall latency time of the network.

**Data collection anonymity:** Privacy of data delivery in wireless sensor networks can be achieved through data aggregation privacy. Data aggregation is concerned with the collection of statistical information about the data gathered by the sensor rather than the data itself, to enhance the utilization of bandwidth and energy [85]. Several publications have targeted the privacy preservation of the data aggregation process.

In [86], a system called 'negative survey' is proposed; it is composed of two protocols: node protocol and base station protocol. The node protocol is used by each node in the network to determine what data will be sent back to the base station. Once the base station receives the data from the sensor nodes, it runs the base station protocol to build the statistical distribution of the data.

Recent work presented in [87] proposes a $(\alpha, k)$ anonymity method based on clustering, to ensure data collection privacy.

**Communication anonymity:** Anonymous communication is concerned with the hiding of communication relationships within the traffic flow, to make an adversary unable to link two communicating parties or link different communications to the same user [88]. Several research efforts have been presented in the literature, to try to protect the anonymity of communications within a WSN.

A scheme called Fortified Anonymous Communication (FAC) deployed temporal privacy and anonymity, to ensure end-to-end location privacy [89]. FAC is claimed in [89] to be able to ensure sender, receiver and link anonymity, source location privacy, base station privacy and energy preservation. The main idea of the scheme is the deployment of an anonymity module, which is responsible for the pre-deployment phase, set-up phase and the communication phase, where security and privacy measures are considered in all these modules to ensure fully anonymous communication. This scheme is believed to be able to withstand local, multi-local and global adversaries.

Another protocol, the Efficient Anonymous Communication (EAC) protocol, is proposed for anonymous communication in [90], where it is said to guarantee the anonymity of the sender, base station and communication relationships. To ensure anonymous communication, four schemes are deployed: anonymous data sending, anonymous data forwarding, anonymous broadcast, and anonymous acknowledgement. Anonymous data sending protects the anonymity of the source node by deploying a global anonymous identity which a source node uses and changes after each message sent. Anonymous data forwarding is concerned with hiding the data forwarding relationship between neighboring sensor nodes. Anonymous broadcasting is used to make an adversary unable to distinguish broadcast messages from other messages, to hide the location of the base station. Anonymous acknowledgement is used to ensure that there is no loss of messages or transmission error within the communication process.

Other work on anonymous communication can be found in [91] which presents an anonymous path routing protocol. The protocol uses data encryption, and anonyms between neighboring sensor nodes and anonyms between the source and destination nodes of a multi-hop communication path. Encryption of data by pair-wise keys is performed to prevent data disclosure by adversaries, and the anonymous communication conceals from adversaries the relation between packets.

### 3.3.1.2    Pseudonomity

Pseudonomity has not been extensively researched in wireless sensor networks, compared to the other privacy goals, as a possible target for a dedicated privacy service. Pseudonomity in wireless sensor networks was used in user authentication [92] [93] and location privacy [94]. In the authentication protocol proposed in [92], the real identity of a user is hidden using a hashing-based random pseudonym. The pseudonym is generated by hashing the identifier of the user and XORing with a random number which is used only once (a nonce). The proposed authentication schemes were shown to resist a number of attacks including login replay attack.

[93] propose a temporal-credential-based mutual authentication and key agreement scheme which includes a user registration phase during which the gateway node generates a pseudo identity. This pseudonym is used during the authentication and key agreement phase.

In the source-location privacy scheme proposed in [94], pseudonyms are generated by using (during the event transmission phase) the identities of the nodes along a route together with a random value, which is shared between each two neighboring nodes in the route. Pairs of nodes, which share a key, use a one-way keyed hash function to create a sequence of pseudonyms by iteratively hashing a random value. The pseudonyms are thus used for identifying the sending and receiving nodes, and for identifying routes (different random values are used for different routes).

### 3.3.1.3    Unlinkability

Many papers have addressed unlinkability in wireless sensor networks. Although not all of these papers have a primary focus on unlinkability, it was achieved as a by-product from solutions to other problems. For example, the main target of [94] was to develop a scheme for source node location privacy. This pseudonym-based scheme relies on creating a cloud of fake packets which take different routes and change appearance at each hop, to makes it hard for an adversary to trace real packets to their source. Privacy protection is achieved by: (i) creating an irregularly shaped cloud of fake packets around the source node; (ii) varying traffic routes; and (iii) using cryptographic operations (such as hash function and symmetric key cryptography) to change the appearance of packets at each hop so as to prevent packet correlation by an adversary.

In the pre-deployment and bootstrapping phase of the scheme proposed in [94], each source node chooses a group of nodes at different number of hops, to assign them as fake source nodes to its packets. The identities of the nodes along a route will be used (during the event transmission phase) with a random value to generate pseudonyms which will be shared between each two neighboring nodes in the route. During the event transmission phase, to protect the location of the source node, the real source node sends events to fake source nodes so that they relay them to the sink, and simultaneously, a cloud of fake packets is activated. The scheme was shown to be resilient to packet-content correlation attacks, time correlation attacks, packet tracing attacks, and packet-replay attacks.

The scheme proposed in [94] achieves unlinkability between pseudonyms because an adversary cannot distinguish event packets from fake packets, this inability makes it infeasible to correlate pseudonyms, as it prevents the adversary from knowing the secret key used in generating the pseudonyms. The proposed scheme also achieves unlinkability between fake and real source nodes because each real source node sends packets to multiple fake sources and each fake source serves multiple real sources. In addition, it is not possible for an adversary to link source node packets and sink packets, thereby the scheme yields source node and sink unlinkability.

Another scheme which indirectly preserved unlinkability in wireless sensor networks is the re-authentication of nodes in a mobile WSN environment [95].

Other papers have proposed different techniques for the implementation of the unlinkability service in WSN-based healthcare systems. Some authors emphasize the importance of unlinkability in their work, however they admit that their work provides weak unlinkability when adversaries perform traffic analysis. An example of these papers is [96] . Other papers (like [97]) rely on random tags to achieve unlinkability; however their scheme does not provide unlinkability in case of traffic analysis attacks because an adversary can simply monitor all traffic from the Personal Digital Assistant (PDA) of the patient, and link all messages to this particular patient. Another scheme for unlinkability in emergency call situations is presented in [98], which proposes a privacy-preserving emergency call scheme designed to ensure unlinkability between transactions and the unique identities of the patients.

Another category of unlinkability techniques in WSN-based healthcare systems is the family of encryption-based techniques, as in [99]. The scheme proposed in [99] relies on encrypting the entire packet (the header, payload, and Message Authentication Code (MAC)). In addition, the protocols change the header, payload and MAC so that the packets appear pseudorandom and cannot be linked by adversaries to the same sender.

### 3.3.1.4    Undectability

There are many mechanisms for preventing an attacker from being able to sufficiently distinguish whether an item of interest (data subject, message, or action, for example) exists or not. In particular, undectability can be achieved by techniques which make the item of interest appear random to all parties except the sender and the recipient(s) [17]. For example, sending encrypted messages in a stream which includes dummy messages (injected to maintain a constant message flow rate) would make data look random for all parties except the sender and the intended recipient(s). Hiding the location of a base station can also be done through a randomness approach [65]. Other approaches for concealing the location of a base station include: hiding the packet destination address (by encrypting the address, packet type and content, for example), de-correlating the packet sending time, and controlling packet sending rates [66]. Similarly, the SPYC protocol uses obfuscation through virtual sensor-region names and cryptographic techniques, to conceal data returned in response to a query [78].

Pfitzman and Hansen [17] stated that dummy traffic can make the number and length of sent or received messages undetectable by everybody except for the recipients or the senders, respectively. Thus, undetectability can be achieved by techniques such as the one reported in [82], which uses dummy messages such that sensor nodes maintain a constant traffic pattern in the network, and makes it hard for an attacker to detect the real source of messages. Similarly, other information flooding schemes can yield undetectability, whereby information about the location of a source node (for example) is concealed through randomized data routing and phantom traffic generation [100], or through a "greedy random walk" (GROW) protocol [101].

For multimedia data, undetectability can be achieved either through a reversible operation or an irreversible operation. Scrambling and encryption are reversible operations for concealing privacy-sensitive information whereas blurring and pixellation are irreversible operations [102]. Steganography is a well-established undetectability mechanism for visual or audio media. For example, one approach removes privacy-sensitive visual information from a video and hides it into the video stream using watermarking [103] [104].

### 3.3.1.5    Unobservability

According to [82], unobservability can be achieved by using a mechanism which combines anonymity with dummy traffic, such that an adversary cannot tell where the real packets are. Consequently, the mechanisms presented earlier under 'source anonymity' which combine fake traffic with anonymity can be considered to be schemes for achieving unobservability. A similar scheme which uses dummy traffic to achieve unobservability can be found in [105]. This technique deploys the dummy traffic concept, to hide real traffic from adversaries, thus defeating global adversaries. In addition, the paper suggests the use of sensors which act as proxies to destroy dummy traffic, in order to decrease the overall cost of extra traffic. Two schemes are suggested for selecting sensor nodes as proxies: proxy-based filtering, and tree-based filtering.

### 3.3.2    Services and mechanisms for location privacy

A brief review of location privacy is presented here, given the importance of being able to hide the location of a healthcare subject, particularly for body-worn wireless sensor networks. Among context-related privacy concerns, location privacy is important in wireless sensor networks, especially given that some healthcare services are linked to the location of the subject, but it may also be the case that disclosure of subject location may breach their privacy. Surprisingly, this tension in requirements has not attracted much attention; to the best of our knowledge, there have been no research papers which have targeted location privacy specifically for WSN-based healthcare systems. Hence, this section provides only pointers to the literature on location privacy in wireless sensor networks at large.

A big research effort has been dedicated to location privacy in general-purpose wireless sensor networks. Location privacy can be categorized into source location privacy [106], sink (receiver/ base station) location privacy [68], and end-to-end data privacy [107]. In [106], techniques for achieving source location privacy were put into 11 categories, namely: random walk; geographic routing; delay; use of dummy or fake data sources; cyclic entrapment; location anonymization; cross-layer routing; separate path routing; network coding; limiting the node detectability; and others. Recent source location privacy techniques are reported in [89], and recent work on sink (base station) location privacy can be found in [108] [68].

Countermeasures against attacks relating to location privacy have been discussed in the sub-section labelled "Base station anonymity" (see Section 3.3.1.1). For example, disguising the location of a base station, to withstand traffic analysis attacks, is reported in [65] [64] [66] [67]. The protection mechanisms are: randomized traffic volumes (multi-parent routing; random walk, random fake paths, fractal propagation) [65]; anonymous topology discovery whereby a pseudo-base station is randomly chosen by the base station to pretend to be a base station and initiate a topology discovery, and fake message injection during the data transmission phase [64]; hidden packet-destination address, de-correlating packet sending time, and controlling packet sending rates [66]; randomized routing paths combined with injection of fake messages into the network [67]; injection of fake messages, sent towards the base station using a biased random walk combined with a routing table perturbation [68].

## 4    TOWARDS HEALTHCARE SYSTEMS WHICH INCORPORATE WIRELESS MULTIMEDIA SENSOR NETWORKS

A wireless multimedia sensor network includes nodes equipped with sensors such as cameras, microphones and other sensors; which produce multimedia content. This content is transferred over the network and/or processed within it. Such networks can be used in a wide range of healthcare applications, such as telemedicine or telecare.

Wireless multimedia sensor networks present a unique set of challenges with regards to privacy protection. One dimension of the challenge is that special privacy protection mechanisms are required for controlling the visibility or audibility of the private information about people which can be contained in images or sounds, particularly if the camera or microphone is placed in private spaces like homes, for example. Another dimension is that the privacy protection mechanisms must operate within bounds imposed by hardware constraints. Visual or aural data demand a high communication bandwidth, and high computational power (with regards to processing and storage), typically resulting in relatively high consumption of electrical energy. Hence, this paper devotes a separate section to wireless multimedia sensor networks.

This section reviews developments towards healthcare systems which include wireless multimedia sensor networks. It includes a review of the associated privacy protection mechanisms and a discussion of practical challenges.

### 4.1 Deployment of wireless multimedia sensor networks in healthcare

Wireless Multimedia Sensor Networks (WMSNs) are networks of sensors which collect different types of digital media, such as audio, video, still images, or other scalar data [109]. Multimedia sensors increase the range of applications for wireless sensor networks [109]. In particular, the deployment of wireless multimedia sensor networks within healthcare systems has been mentioned as one of their many possible applications [110] [109].

However, only a few papers have attempted to offer actual solutions for healthcare systems. For example, [111] focuses on overcoming the challenge of the high bandwidth required in live telemedicine applications which use video and audio streaming. The in-house healthcare application scenario for wireless multimedia sensor networks, which is illustrated in Figure 11, is based on scenarios presented in [15] and [34]. They outline how different sensors can be deployed to monitor patients. The figure shows a possible view of an apartment where patients with different needs (for example: chronically ill patients, handicapped or elderly people) may live. Each patient would wear or have implanted the appropriate sensors required to monitor their health or general physical or mental state. For example, oxygen saturation, heart rate, body temperature, and blood pressure sensors can be used to monitor a chronically-ill patient [15]. Other sensors can also be used to monitor the environment surrounding the patients. They may include humidity, temperature or pressure sensors, and Radio-Frequency IDentification (RFID) sensors.

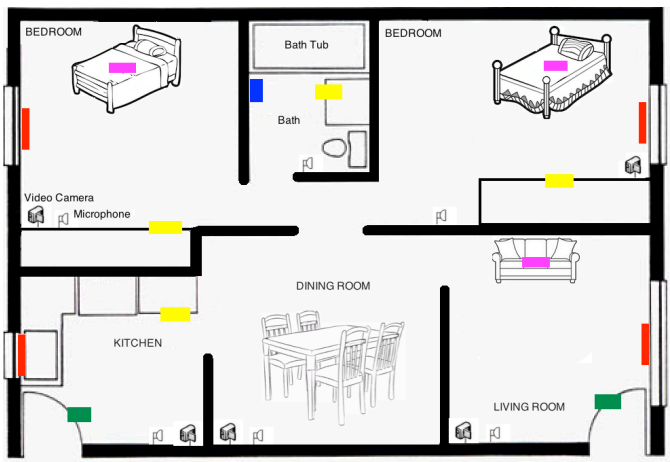

Figure 11. A view of a house equipped with a wireless multimedia sensor network, for healthcare monitoring. The color coding used for sensors is: green for door sensors, red for window sensors, pink for pressure sensors, blue for humidity sensors, and yellow for RFID sensors. In addition to the sensors shown in the figure, other sensors could be worn by, or be implanted in, patients who are in the house.

As depicted in Figure 11, door sensors and window sensors can be used to detect whether doors or windows are opened or closed. Pressure sensors can be attached to sofas or beds to detect if someone is sitting or lying on them. Humidity sensors can be used in bathrooms to monitor the humidity level. RFID sensors can be attached to household items such as a medicine cabinet, refrigerator, bedroom closet, or other items, to detect when a patient has used them. Microphones and video cameras can be used to monitor the patient, for example to hear their requests for help, or monitor their gait or posture to detect if they stumble into an obstacle or fall down.

A seamless healthcare environment can be built based on a wireless multimedia sensor network, configured as a multi-tier architecture (see Section 2), with sensors that collect different types of data (scalar, audio and video). Once the sensors capture data, they could be sent to a nearby data manager or gateway (for example, a computer or a mobile phone). The data manager could then send the data to a base station for relaying to a remote server for further analysis or storage, and possibly to a caregiver. In case of emergency, such as abnormal fluctuations in the readings of the sensors or a patient falling down, alerts can be sent to a caregiver for action, or automated phone calls can be made, asking for intervention by other healthcare professionals, ambulances, hospitals or emergency centers.

### 4.2 Privacy preservation in wireless multimedia sensor networks for healthcare

Distributed multimedia sensor networks, often implemented as wireless multimedia sensor networks, have been used for monitoring sick patients [112] and in other healthcare contexts. In some situations, multimedia analytics automatically extract, from the visual or audio data collected by the wireless multimedia sensor network, high-level information relevant for the monitoring task. The original visual or audio data can then be discarded, together with privacy-sensitive information contained therein, such that only high-level privacy-preserving information is sent to healthcare staff. In many situations however, due to technical challenges which limit the capability of multimedia analytics, the visual or audio data is presented unprocessed to a human observer, for decision and action, in critical conditions or otherwise. Hence, various methods have been proposed in the literature, to conceal from the human observer as much of the personally identifiable information as circumstances require.

#### 4.2.1 Privacy protection mechanisms

Due to the potentially sensitive nature of images or sounds, which may reveal private information about the people being monitored, privacy services are an important prerequisite towards the acceptance of wireless multimedia sensor networks in healthcare, by patients, healthcare providers, policy makers, and regulators [113]. Critical information about people, which may be revealed by wireless multimedia sensor networks, is not only their identities but also their behavioral patterns. Thus, wireless multimedia sensor networks require special privacy protection mechanisms, to preserve the privacy of patients by addressing the threats associated with the use of multimedia data [114].

These mechanisms must comply with the constraints of wireless multimedia sensor networks. The deployment of such networks in healthcare systems is constrained by many factors such as the demand for high bandwidth, fast processing and quality of service in wireless multimedia sensor networks, and their relatively high consumption of electrical energy [109]. Monetary cost also imposes further constraints when designing a privacy-aware wireless multimedia sensor network [16]. Consequently, under the constraints on healthcare application environments which prevail at the time of writing, typically only important or light-weight privacy services are considered and less important ones discarded or made optional.

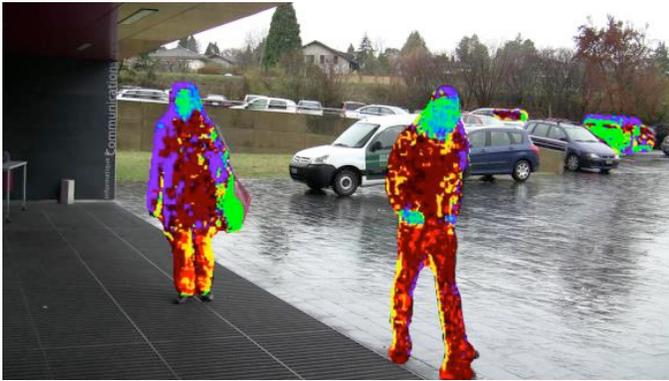

Figure 12. Obfuscation of personally identifiable information in an image [115]

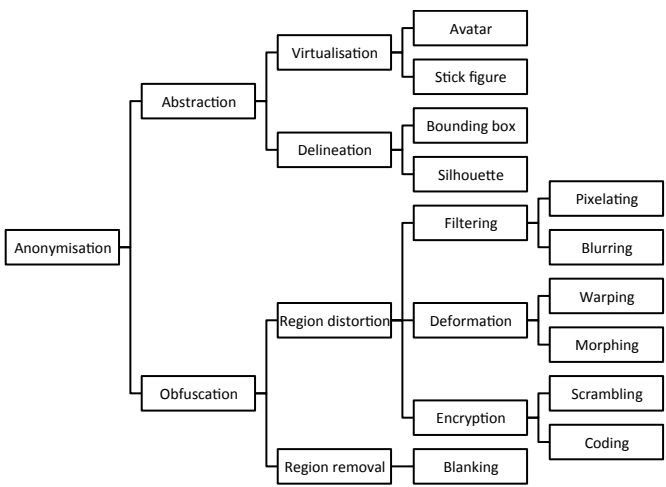

Figure 13. Classification of anonymization techniques for visual data.

Privacy protection for image or video data is often achieved by *anonymization* of the data, which is typically done through a selective protection of image regions, by abstracting or obfuscating the personally identifiable information in the region (see Figure 12). Identity concealment is often applied to specific parts of the visual data, specifically those which correspond to visually meaningful entities like a person or identifiable parts of their body (such as the face). Figure 13 presents a classification of techniques for anonymizing visual data.

Abstraction of an image region, which contains privacy sensitive data, can be done by replacing the region by a virtual entity such as an avatar or stick figure. Alternatively, only an outline of the region can be displayed (as a silhouette [116] or a bounding box, for example; with the inside of the region being masked out). The region delineation operation is often performed in the spatial domain. An example of data abstraction is provided in the UbiSense Distributed Multimedia Sensor Network [117], which was developed for automated homecare monitoring of the elderly, built on gait and posture recognition. In the UbiSense solution, video sensors are installed in the environment and they are used together with body sensors and radio-frequency identification (RFID) devices. Video is processed at the camera, to retain only abstract information about the monitored scene. The abstract information includes shape and other relevant information necessary for detecting anomalies in the gait or posture of the elderly. Privacy is enhanced because only this abstracted information is transmitted and processed within the network. The weakness of the scheme is that it may still be possible to identify individuals using the retained information because gait and posture contain some characteristics which are sufficiently unique to each individual. Hence, the person could be identified from the retained information.

Similarly, in the networked sensor tapestry (NeST) architecture [118], the raw sensor data is processed to remove personally identifiable information and the resulting data are communicated to other nodes in the network. Fidaleo et al. [118]—offered NeST to facilitate rapid prototyping and deployment of wireless multimedia sensor networks for surveillance applications. The architecture was designed for the secure capture, processing, distribution and archiving of multimedia data. It allows control over the privacy of multimedia data captured by the system. One component of the architecture is the privacy buffer, which prevents access to private information, or transforms private information by removing personally identifiable information, such that only the behavior of an individual under surveillance is conveyed but not their identity. The selection of the private data which needs to be obfuscated is achieved by privacy filters. The resulting obfuscated data are transferred over a network which deploys the secure socket layer protocol and client authorization, for network-level protection.

Obfuscation can take the form of removal or distortion of the relevant image region(s). Region blanking is a common removal operation. For example, in [112], image processing selectively masks subjects, and the resulting video data are fused with information from other sensors. Subjects wear RFID tags which specify whether the privacy of the individual needs to be preserved; this triggers or disables the masking.

Another popular image distortion operation is image filtering, implemented as region pixelization or blurring. Filtering can be performed in the spatial, temporal or spectral domain [119] [120] [112] [121] [102].

A region can also be distorted by deforming it through an appropriate warping technique [122] or morphing technique [123]. Warping is purely a geometric transformation, as it changes only the shapes displayed in the affected region, whereas morphing can change shapes and pixel colors or region texture.

Alternatively, encryption is a common approach for distorting an image region, through scrambling or coding the content of the region. Encryption can be implemented in the spatial domain [124][125], or temporal or spectral domain. For instance, it can be applied to a region of interest in compressed video by trimming or scrambling the code blocks for the given region, for example by pseudo-random inversion of the sign of selected transform coefficients (during encoding) or pseudo-random permutation of code bits in the code stream [126] [127] [116] [128]. Scrambling by sign inversion or bit permutation offer the advantages of low computational cost and little effect on coding efficiency. A physics-based scrambling approach was developed by Sedky, Chibelushi and Moniri [115]. They used an image formation model firstly to separate from the background scene the pixels corresponding to human subjects appearing in the foreground of a video. Secondly, they extracted the spectral reflectance of the surface material corresponding to each of those pixels. They then replaced each of the extracted pixels by the color corresponding to the wavelength which has the lowest amplitude along the reflectance curve.

The operation to conceal privacy-sensitive information can be lossless (reversible) or lossy (irreversible) [102]. For example, scrambling and encryption are lossless operations, whereas blurring and pixelization are lossy. One approach for secure reversible data hiding consists of removing privacy-sensitive visual information from the video and hiding it into the video stream using watermarking [103] [104].

Privacy requirements vary with the application context; hence, the choice of appropriate technique will be context-dependent, and possibly be influenced by the level performance of the technique. Some performance assessment studies have been reported in the privacy protection literature. Gross et al. [121], and Dufaux and Ebrahimi [102] showed that pixelization and blurring achieved relatively low privacy protection against face recognition algorithms, and they observed that scrambling yielded better protection. In another comparative performance evaluation, [129] assessed five privacy protection techniques (namely: Gaussian blurring, pixelization, masking, warping, and morphing) by investigating the influence of each technique on the performance of each of three face recognition algorithms (respectively based on principal component analysis, linear discriminant analysis, and local binary patterns). Among other findings, the results showed that morphing was the best choice among the evaluated privacy filters.

In view of the significant amount of privacy sensitive information potentially conveyed by visual data, most of the research on privacy protection in distributed multimedia sensor networks has focused on privacy in vision-rich systems. However, obfuscation can also be applied to audio. For example, audio distortion can be applied by shifting the pitch of the audio signal, to conceal identity-bearing information carried by voice [130] [131] [116] [102]. Furthermore, multimedia data can be scrambled by applying a random permutation, or shuffling, to spatial, temporal or spectral characteristics of audio or video data.

Winkler and Rinner [132] classify identity-bearing information into *primary identifiers* and *secondary identifiers*. Primary identifiers relate directly to an *individual* (examples of such identifiers are the face, gender, race, gait of the person, or items or devices carried by the person – such as a mobile phone which is carried by a patient). Hence, primary identifiers can be used to identify the person directly. Secondary identifiers are those which relate to the *environment* surrounding a person and to *interactions* between the person and their environment. In particular, with reference to secondary identifiers, information which can indirectly lead to the identification of an individual are the *where* (location), *when* (time) and *what* (actions) [133]. Although Winkler and Rinner [132] introduced the concepts of primary identifiers and secondary identifiers when discussing visual information, the concepts extend to other information. In particular, secondary identifiers are linked to the context-related privacy concerns and services, which were mentioned in earlier sections (see Sections 1.4 and 3).

Although, the protection of secondary identifiers can be complex, some relatively simple solutions have been proposed. For example, obfuscation can also be applied to identifying information associated with communicating devices, such as those carried by a patient. Furthermore, the reader is referred to Section 3.3, which discusses many relevant techniques.

Besides achieving anonymity, a derived benefit of removing identifiers from the data is that anonymization can make the link between related entities ambiguous, and hence it can enhance unlinkability. Similarly, encryption can be used to make data entities unlinkable, as a result of the apparent randomness of the data which it produces.

Undetectability is another privacy goal often pursued for visual data in wireless sensor networks. Kundur and co-workers proposed a distributed visual secret sharing technique, to counter eavesdropping (a passive attack) by providing a level of confidentiality and undetectability suitable for applications such as geriatric monitoring [46] [134] [135]. It is anchored on a decentralized scheme to achieve confidentiality and undetectability for dense visual sensor networks.

A distributed privacy paradigm was proposed for vision-rich sensor networking in [46]. It is based on the control of dynamical systems, which takes the form of a distributed visual secret sharing paradigm [134]. The paradigm is detailed in [134], and it was applied in two different ways in [134] and [135]. It enhances confidentiality and undetectability in a distributed manner. Cameras are grouped such that each captures visual readings which are correlated with readings from other cameras in the group, because the group readings pertain to the same scene and the camera readings are taken over the same time interval.

Data from a camera is used to generate a share (of the secret scene) which is transmitted to the base station along multi-hop paths (pairwise encryption, with a pairwise key, is applied to the share transmitted between two adjacent nodes) which are disjoint from the paths of the other cameras. Only the base station is likely to be able to reconstruct the secret (a composite secret) by aggregating the shares that it receives (the more complete the set of uncorrupted shares received, the better the reconstruction), given that most of the shares will reach it. On the other hand, a corrupt node (captured by an attacker) has access to only a small fraction of shares, such that it is not able to reconstruct the secret. The quality of the image approximation reconstructed by the base station increases proportionally to the number of shares which reach the base station uncorrupted (see Figure 14). Conversely, the quality of the image reconstructed by an eavesdropper degrades proportionally to reduction in the number of shares captured by the eavesdropper.

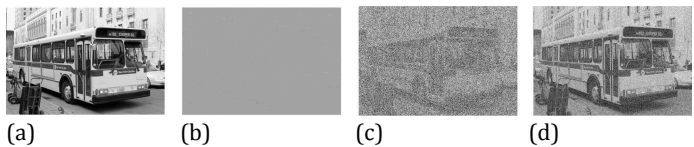

(a)          (b)          (c)          (d)

Figure 14. Illustration of the increasing quality of the image approximation reconstructed from an increasing number of shares of a visual secret. (a) Original secret image, (b) sample share, (c) image produced by combining 10 shares, (d) image produced by combining 40 shares. (Reproduced from [46])

Relaxing the quality of the regenerated image is exploited to produce a lightweight solution suitable for low-cost sensors, by reducing the complexity of pixel processing computations and the size of the shares used for visual secret-sharing, thereby reducing processing, storage and bandwidth complexity, which are important considerations in distributed multimedia sensor networks.

### 4.2.2    Practical challenges

Wireless multimedia sensor networks have to contend with many practical challenges. Wireless multimedia sensor networks have to contend with the fact that multimedia information (such as digital images, video, and audio) often requires a high communication bandwidth, and is associated with high storage cost and possibly high computation cost if in-network processing of digital media is required. Sensor networks often have constraints imposed on the processing and storage capability of their nodes, and on the available electrical power. Thus, wireless multimedia sensor networks often face significant technical challenges to deliver adequate speed for the communication between nodes, or to provide the required processing or storage capacity within the network, while minimizing electrical power consumption. For instance, encryption of visual data can be expensive in terms of processing, storage and transmission cost.

Fortunately, the storage and transmission capabilities of sensor nodes keep increasing; wireless sensor networks available today often have local processing and storage capability at the nodes. Local processing and storage may reduce the amount of data transmitted over the network. This is because processing data at the node may extract only the relevant information from the raw data. Then, the raw data may be discarded, or data needed by the node but not required by any other node may be stored locally. Examples of local processing include data compression, or extraction of abstract data such as silhouettes, which discard or reduce identity-bearing information. An illustrative example of local processing was implemented in the TrustEye project [136]. The image sensing unit in TrustEye implements privacy protection as one of its inherent features. The unit analyses the content of the captured image or video, and only releases non-sensitive data (such as anonymized data, together with statistical or other abstracted data) to the rest of the visual sensor network.

The caveat of the local processing approach is that processing will require expenditure of electrical energy at the node, and hence processing may be restricted to light-weight algorithms, or to limited data volumes, in some application contexts. Indeed, wireless sensor networks in the healthcare context would fairly often include small, distributed body-worn or implanted sensor nodes, and possibly some video and audio sensor nodes transmitting or relaying health-related data to a central observation station. Some of these nodes would be battery-powered and are currently often constrained by the limited power available for signal processing tasks such as cryptography, and for data storage and transmission. Resource limitations often impact on technological or economic feasibility. Where lightweight privacy protection techniques are a possibility, the reduction in the computational demand of the solution needs to be weighed against real-time performance requirements which may be adversely affected by light-weight techniques.

To address the technical challenge of achieving sufficient communication speeds for multimedia data, while conserving electrical power, some researchers have investigated sensor networks with nodes which use free-space optical transmission of broadband data, coupled to light-weight secure protocols, to transfer data using significantly less energy than omnidirectional radio-frequency transmission [46]. Free-space optical transmission delivers wireless line-of-sight communication technology, using nodes which include signal transmission and reception hardware, which respectively comprise a light emitter and a photodetector. The light emitter (such as a laser diode or a light-emitting diode) transmits light beam signals through the air, such that the beam reaches the photodetector (a photodiode, for example).

Free-space optical transmission offers ultra-high communication bandwidth. Compared to radio-frequency transmission, it also offers lower power consumption, less regulatory restrictions on what part of the frequency spectrum is used for communication, and a higher difficulty for eavesdropping owing to the compactness of the laser beam used for communication in comparison to omnidirectional radio-frequency transmission. However, the bit rate drops or the error rate increases or the reception is interrupted in conditions which adversely affect (by signal attenuation or interference or occlusion) the transmission or reception of the optical beam. These conditions include rain, snow, and fog or intense sunlight, or physical obstruction due to objects or the uneven terrain (if outdoors) where the WSN is deployed.

In the context of visual sensor networks, Winkler and Rinner [132] argue that a trustworthy sensing unit must address the challenges of: protecting identity-bearing information; trading-off privacy against the utility of the system (for example, a balance has to be struck between the degree of anonymization of the visual data and the requirement to retain information which is adequate for monitoring behaviors effectively); meeting the constraints of resource-limited sensing devices; correlating securely the original image data with the higher-level data (such as the detected behavioral events) generated by the user application software. Furthermore, some applications may require that the original image data be linked securely to information generated at the application level (such as detected events which are relevant to behavioral information about a patient that is extracted from video content).

## 5  PRIVACY ATTACKS, THREAT ANALYSIS, AND ASSESSMENT OF PRIVACY PROTECTION IN WIRELESS SENSOR NETWORKS

Wireless sensor networks are beset by vulnerabilities (such as vulnerability to physical tampering) which are associated with physical devices or their physical operation, and by vulnerabilities related to software (such as the operating system, the implementation of communication protocols, or the application software). An attacker may exploit the vulnerabilities, and access private health information, by directing attacks at the physical devices, the communication channel, or at the content of the actual data. Hence, rigorous design and assessment of privacy-preserving services are crucial ingredients for the development of well-engineered wireless sensor networks which are appropriate for deployment in real-world healthcare systems.

This section focuses on analysis methods for privacy threats, and on assessment of resilience to privacy attacks, in wireless sensor networks. It presents a review of threat analysis and assessment methodologies which could be deployed to identify privacy protection services required in wireless sensor networks for healthcare, or to gauge the vulnerabilities of an existing system or prospective system, and determine its level of privacy protection, based on the strength of the attack mitigation techniques which are already deployed in the system. Furthermore, the section includes an overview of privacy attacks and related countermeasures, as a prelude to threat analysis and assessment.

### 5.1  Privacy threat analysis

Ideally, privacy violations should be prevented before they occur. The ability to spot privacy threats, before an actual attack is perpetrated, is therefore highly desirable. Attacks and threats are intimately linked (given that an attack realizes a threat). Thus, this section begins with a review of attacks and their countermeasures, followed by a review of methodologies for privacy threat analysis.

### 5.1.1  Attacks and countermeasures

#### 5.1.1.1  Breach of security goals

In wireless sensor networks, a sensor node or other type of node can be a physically small and portable device; and they might be deployed without mechanisms to protect them from falling physically into the wrong hands. It is thus possible for such a device to be seized by an attacker, or for the attacker to deploy fake nodes [137]. Such an attacker may launch active attacks, whereby the attacker tampers with the device or data, to compromise the integrity of the latter. These attacks are normally tackled through security countermeasures.

In addition, wireless communication brings the possibility of attacks targeted at the vulnerabilities of radio communication, given that nodes transmit to each other over the air. Hence, active attacks may also be launched through injecting fake data or modifying data [25]. An attack may also be passive. For example, in an eavesdropping attack [105], an adversary can gain access to network traffic by listening to the communication.

Kotz [138] proposed a threat taxonomy for the broader area of mobile health (mHealth) privacy (which can be viewed as a superset for privacy in wireless sensor networks). The taxonomy singled out the misuse of patient identifiers, unauthorized access, and modification or disclosure of personal health information, as main threat types. Although Kotz [138] did not discuss countermeasures for attacks linked to these threats, a wide range of countermeasures have been investigated extensively in the security literature.

It is indeed acknowledged generally that privacy protection also requires security countermeasures to deal with attacks. For example the threat of unauthorized access to personal health information can be addressed through measures such as: access control for people and devices (e.g. based on authentication protocols and mechanisms); consent management; and auditing. Such attacks may lead to privacy breaches.

Security attacks and their corresponding countermeasures have already received ample coverage in the security literature, and the relevant literature is quite broad. Therefore, this paper does not cover them. The coverage in this paper of attacks and countermeasures focuses on privacy-specific attacks, to avoid making the paper inordinately long. The interested reader is referred to literature surveys and taxonomies, such as those published in [18] [19] [20] [21] [22] [23] [24] [5]. These papers review security attacks and countermeasures for attacks such as sink-hole attack, worm-hole attack, grey-hole attack, Sybil attack, hello-flood attack, selective-forwarding attack, denial-of-service attack, snooping attack, modification attack, routing-loop attack, and masquerading attack. In addition, [7] [25] discuss a number of security countermeasures for wireless body sensor networks.

### 5.1.1.2    *Breach of privacy goals*

An eavesdropping attack (discussed in the previous section) could lead to gathering actual private data, but it can also be a precursor for further attacks. These attacks may take the form of analyzing the communication patterns between network nodes (traffic analysis attack), or analyzing the content of the data transmitted between the nodes (content analysis attack, also known as data analysis attack) [26].

In the taxonomy produced by Li et al. [26], for privacy-preserving techniques in wireless sensor networks, data-oriented privacy concerns (which focus on the privacy of the actual data or data queries) are distinguished from context-oriented privacy concerns (which focus on contextual information such as location and timing information). These two types of privacy concerns may be infringed by data analysis attacks and by traffic analysis attacks, respectively [26]. TABLE II. summarizes countermeasures against privacy attacks, with emphasis on data analysis attacks, whereas TABLE III. summarizes countermeasures against traffic analysis attacks, and Figure 15 depicts a classification tree of both types of attacks and threats, and privacy protection mechanisms.

### 5.1.1.2.1    Data analysis attacks

According to Li et al. [26], data analysis attacks in wireless sensor networks can be launched by an external adversary (who eavesdrops on data communication between nodes) or by an internal adversary (such as a node of the network which has been tampered with by an attacker, and is under the control of the attacker), and they may target data aggregation processes, or data query processes. Encryption and authentication are established protection countermeasures against data-analysis attacks by an external adversary, to achieve the security goals of confidentiality and integrity, which also indirectly yield privacy protection. However, these security countermeasures might not be effective against an internal adversary, who might seize secret encryption information (the encryption keys, for example) or authentication information.

Encryption could be successful as a possible countermeasure against such an internal adversary, if it is implemented as end-to-end encryption, between the data source and the base station, such that no intermediate node can decrypt and access the data. This countermeasure comes at the cost of increased traffic and consequently increased energy consumption, because intermediate nodes cannot perform data aggregation to reduce traffic volume (by using operations like Sum, Min, Max or Median aggregation). Hence, hop-by-hop encryption is often preferred, but it requires additional protection mechanisms (such as privacy-preserving data aggregation) against internal attacks wherein the encryption keys are in the hands of the attacker.

Privacy-preserving data aggregation techniques typically apply a perturbation to the data. For example, random noise can be injected into the data (as done in cluster-based private data aggregation (CPDA) [139], or each data item can be sliced and the resulting slices shuffled randomly (as done in Slice-Mix-AggRegaTe (SMART) [139]. In the CPDA algorithm, the noise is designed for cooperative sensor nodes, such that precise aggregated values can be recovered by the aggregator. However, the CPDA algorithm has high overheads (communication and computation), and collusion by a number of sensor nodes exceeding a set threshold can allow these nodes to collaboratively reveal the data of some other nodes [140]. The SMART algorithm offers low computation overhead, but it increases the number of messages in the network, with a consequent increase in energy consumption. SMART is also susceptible to collusion by a number of sensor nodes above a set threshold [140].

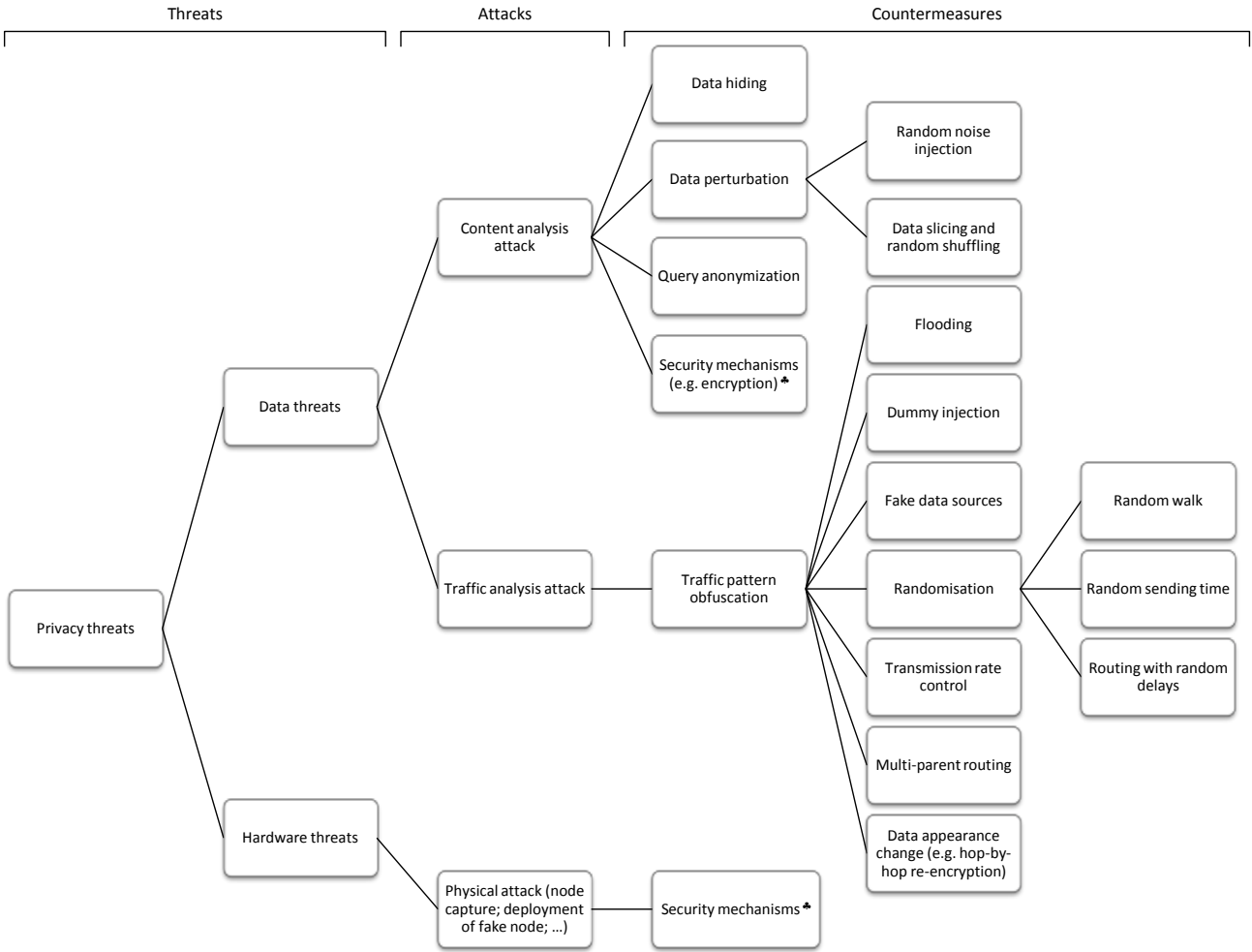

Figure 15. Privacy attacks and protection mechanisms in wireless sensor networks. ⋅Readers interested in security attacks and countermeasures are referred to the security literature such as [18] [19] [20] [21] [22] [23] [24] [5] [7] [25]

Countermeasures against data analysis attacks which target data queries include data hiding techniques [141] and anonymization techniques [142]. In data hiding techniques, the response to a query includes the data requested by the query and some other dummy data from the network. The scheme attempts to prevent the attacker from knowing the query through a deduction process based on data returned in response to a data query. The downside of such data hiding techniques is that enlarging the volume of returned data increases traffic and thus results in increased energy consumption. Anonymization techniques hide the identity of the source of the query. For example, in [142], the identity of the source is hidden by associating the query with tokens obtained by the source, in place of the identity of the source.

5.1.1.2.2    Traffic analysis attacks

A traffic analysis attack may encompass communication patterns over the whole network (it is then called a global attack), or focus on a localized part of the network (it is then called a local attack, which can take the form of a hop-by-hop trace attack, for instance). As wireless sensor networks use hop-by-hop transmission of data, analysis of the traffic pattern between hops can reveal which node is the base station, or which node is the source of the data. Such analysis can also reveal what time the data was captured by a sensor. For example, a hop-by-hop-trace (or path-tracing) attack [105] [106] may be deployed, whereby an adversary trails the traffic from node to node, so as to get to the source or sink node, which is the origin or final destination of the traffic. An adversary may also deploy a timing analysis attack [143], by monitoring the pattern of the timing of transmissions, and hence infer information such as the structure of the network or the correlation between incoming and outgoing traffic.

TABLE II.    SUMMARY OF PRIVACY-PROTECTION MECHANISMS AGAINST ATTACKS, WITH EMPHASIS ON ATTACKS RELATED TO THE CONTENT OF THE DATA WHICH IS TRANSFERRED OVER THE NETWORK

| Attacks | | Privacy protection mechanism | Privacy goal | | | | | Notes |
|---|---|---|---|---|---|---|---|---|
| | | | Anonymity | Pseudonymity | Unlinkability | Undetectability | Unobservability | |
| Physical attack | Active attack | ❖ Seizure of node ❖ Deployment of fake node | Privacy protection as a by-product of security countermeasures | | | | | Security defenses [137] based on tamper-resistant packaging, or redundancy (e.g. routing protocols which sends every packet along multiple paths and checks the consistency among the packets received at the destination). |
| Non-physical attack | Passive attack | ❖ Eavesdropping attack | Privacy protection as a by-product of security countermeasures | | √ | √ | | Encryption is a popular defense [137]. Hardware solutions are also possible. For example, directional antennas can reduce the eavesdropping probability. |
| | | ❖ Content analysis attack: • targeted at the actual data | Privacy protection as a by-product of security countermeasures [137]: *Encryption*: | | | | | |
| | | | • End-to-end encryption | | √ | √ | | Can be effective against an internal adversary, at the cost of increased traffic and hence increased energy consumption. |
| | | | • Hop-by-hop encryption | | √ | √ | | Can be effective against an external adversary. |
| | | | *Authentication* | √ | √ | √ | | Authentication can filter out malicious network nodes (or users). Mechanisms include cryptography-based methods (such as pairwise or group-wise node authentication), and anonymous authentication methods. |
| | | • targeted at data aggregation | *Data perturbation*: | | | | | |
| | | | • Injection of random noise into the data (Cluster-based Private Data Aggregation (CPDA) [139] | | √ | √ | | CPDA allows the aggregator to recover precise aggregated values, but it has high overheads (communication and computation) and is susceptible to collusion by nodes. |
| | | | • Data slicing and random shuffling of slices (Slice-Mix-AggRegaTe (SMART) [139] | | √ | √ | | SMART offers low computation overhead, but has communication overheads. It is also susceptible to collusion by nodes. |
| | | • targeted at the data query | *Data hiding* [141] | | | √ | | The response to a query includes the data requested by the query and some dummy data, at the cost of increased traffic. |
| | | | *Anonymization* [142] | √ | | | | √ | The identity of the source of the query is hidden (tokens obtained by the source are used in place of the identity of the source). This method can be combined with dummy tuples. |
| | | ❖ Traffic analysis attack | See TABLE III. | See TABLE III. | | | | | - |

TABLE III.    SUMMARY OF PRIVACY-PROTECTION MECHANISMS AGAINST ATTACKS WHICH RELATE TO NETWORK TRAFFIC

| Privacy goal | | Privacy protection mechanism | Notes |
|---|---|---|---|
| Anonymity | Base station anonymity | Base station relocation [61] | Depends on whether motion is possible. Inhibits long-time traffic analysis. |
| | | Network nodes transmission power / range increase [62] | Raising the transmission power of nodes increases the correlation between neighboring nodes and makes traffic analysis more difficult. |
| | | Distributed beamforming [63] | Beamforming boosts base station anonymity; low communication overhead. |
| | | Randomized traffic volumes [65]: | Randomized traffic volumes protect the location of the base station. They defend against traffic analysis. |
| | | • multi-parent routing [65] | Randomly selecting one of the parent nodes, to forward data to the base station, inhibits traffic analysis attacks by making it hard to detect a traffic pattern which may lead to the base station. |
| | | • random walk [65] | A node forwards packets to its parent nodes, based on a random forwarding algorithm, thus distributing the traffic of packets and decreasing the effectiveness of traffic analysis attacks (rate monitoring attacks). |
| | | • random fake paths [65] | Fake routes are introduced from the node to the base station, to reduce the effectiveness of traffic analysis attacks such as time correlation attacks. |
| | | • fractal propagation [65] | Fake messages are created and propagated in the network, to yield areas of high activity and randomness in the communication pattern, and thus defend against traffic analysis attacks like the rate monitoring attacks. |
| | | Uniform traffic volume [66] | Controlling packet sending rates creates a uniform traffic volume in the network, which makes it difficult for a traffic analysis attack to deduce the location of the base station by using measurements of traffic volume. |
| | | Data appearance change through encryption [66] [94] | Encryption (of packet destination address, packet type and content, for example) can inhibit traffic analysis attacks such as packet correlation attacks. |
| | | Random sending time [66] | Randomizing the packet sending times introduces temporal variation which reduces the effectiveness of time correlation attacks. |
| | | Randomization of routing paths and injection of fake messages [67] | Randomizing routing paths towards the base station and injecting fake messages, to uniformly distribute the incoming and outgoing traffic at a sensor node, inhibits traffic analysis attacks such as packet tracing attacks. |
| | | Fake message injection, and biased random walk with routing table perturbation [68] | Fake packets and random paths can act as countermeasures against traffic analysis attacks by a local adversary. |
| | | Anonymous topology discovery through a randomly-chosen pseudo-base station, together with fake message injection during the data transmission phase [64] | Randomly-chosen pseudo-base stations, together with the injection of fake packets, inhibit traffic analysis attacks such as packet tracing attacks; the countermeasures introduce many fake paths. The scheme is combined with a simple random walk algorithm, to hide the location of the base station. |
| | User anonymity | Anonymous two-factor authentication [69] | User authentication based on a smart-card and a password. |
| | | Smart CArd based user authentication scheme for WSN (SCA-WSN) [72] | Lightweight anonymous two-factor user authentication which requires a smart card. |
| | | Biometrics-based authentication [75] | Computationally efficient anonymous user authentication based on biometrics. |
| | Query anonymity | SPYC protocol (query obfuscation using virtual region names and cryptographic techniques) [78] | The interaction of the client with the sensor network comprises: building a virtual naming space for sensors, and accessing sensor readings through region-based source routing which is based on virtual region names. |
| | | Distributed Privacy Preserving Access Control (DP$^2$AC) [79] | Anonymous access to sensor data using validated tokens bought by the client from the network owner, whereby token generation involves blind signatures. |
| | Source anonymity | Dummy messages for constant network traffic pattern [82] | A constant traffic pattern and hiding real packets within slots which include dummy packets inhibit traffic analysis attack from detecting the data source. |
| | Data collection anonymity | 'Negative survey' system [86] | Data aggregation privacy is preserved through two protocols: each node determines what data to send to the base station (node protocol); the base station builds a statistical distribution of node data (base station protocol). |
| | | (α,k) anonymity method [87] | Privacy-preserving data collection model which is based on clustering. |
| | Communication anonymity | Fortified Anonymous Communication (FAC) [89] | An anonymity module is deployed to ensure fully anonymous communication, for end-to-end location privacy. |
| | | Efficient Anonymous Communication (EAC) protocol [90] | The protocol can withstand local, multi-local and global adversaries by protecting sender anonymity, base station and communication relationships. |

TABLE III. (Continued from the previous page)

| Privacy goal | Privacy protection mechanism | Notes |
|---|---|---|
| Pseudonymity | Hashing-based random pseudonyms [92] | User identity is hidden using hash-based random pseudonyms. User authentication is shown to resist attacks including login replay attack. |
| | Pairwise generation of sequence of pseudonyms [94] | Pairs of nodes use a one-way keyed hash function to create a sequence of pseudonyms, used for identifying the sending and receiving nodes, and for identifying routes. |
| Unlinkability | Combination of: cloud of fake packets, varying traffic routes, and data appearance change at each hop through encryption [94] | The combination of fake packets, varying traffic routes, and data appearance change, and pseudonyms is shown to be resilient to packet-content correlation attacks, time correlation attacks, packet tracing attacks, and packet-replay attacks. |
| Undetectability | Randomness (e.g. through encryption, dummy packets, random sending time) [65] [66] [78] or steganography [103] [104] | The item of interest is concealed by making it appear random, or through obfuscation or steganography. |
| Unobservability | Combination of anonymity with dummy traffic [82] [105] | Unobservability is achieved by making a global adversary unable to tell where the real packets are, through a mechanism which combines anonymity with dummy traffic. In addition, sensors which act as proxies can destroy dummy traffic, to decrease the cost of extra traffic |

In another form of traffic analysis attack, known as rate monitoring attack, the adversary watches the transmission rate of nodes, to identify nodes with a high transmission rate, as an indication of the probability that the nodes are nearer to the source or to the sink. A time correlation attack [105] is yet another traffic analysis attack, whereby the adversary records transmission times between a node and its neighbors, measures the correlation between these times and uses this information to detect which neighboring node(s) forward(s) the packets from the current node, and thus estimate the route followed by a packet travelling to the sink. Kûr and Stetsko [144] classified traffic analysis attacks in wireless sensor networks into three main types: rate monitoring attacks, time correlation attacks, and content analysis attacks. In the content analysis attack, an adversary gleans the relevant information (such as the location of a base station) from packet headers and payloads.

Besides the above attacks, an adversary can monitor transmission patterns or the transmitted signals to extract other indicators, such as: the angle of arrival [145] which would allow the adversary to establish the direction of a transmitting node, or the received signal strength [145] from which the distance between the node and the adversary can be estimated. Furthermore, an identity-analysis attack [105] can be deployed. In a similar vein to the time correlation attack, the adversary (in an identity-analysis attack) tries to establish the relation between nodes by measuring the frequency of use of identities.

It is important to note that traffic analysis can enable an attacker to infer private information without the need to decrypt the data. Hence, encryption may not be effective against a traffic analysis attack, despite being a good countermeasure against eavesdropping and injection or modification of data [25] [60]. For example, [33] includes a description of the 'fingerprint and timing-based snooping' (FATS) attack whereby an attacker eavesdrops on radio transmissions, analyses the data transmitted over the network and deduces private information from it, with no decryption required. The attacker uses transmission time and unique features (referred to as the "fingerprint") of the radio frequency waveform of each transmitter. The attacker gathers the time-stamps and fingerprints of radio transmissions, to associate each message with a unique transmitter. The attacker performs further analysis to infer the location of the subjects (such as: dining room, kitchen, bathroom, …, for example) and the type(s) of sensor (motion sensor, heart-rate monitor, …, for example) worn by them. This information is in turn used to infer the activities of the subjects, and therefrom their health conditions. Countermeasures for a FATS attack include signal attenuation (to reduce the effectiveness of eavesdropping), and perturbation of the communication (through fake message, delays, ...).

TABLE III. summarizes countermeasures against traffic analysis attacks. A more detailed discussion of countermeasures is given in Section 3.3.

### 5.1.2 Threat analysis methodologies

"Privacy-by-design" is a privacy engineering approach which is used to elicit (during the design phase) the privacy threats for a software system, to avoid the challenge of implementing privacy services as a system component bolted-on after the software engineering process. Privacy threat models can be used to discover the privacy requirements of the system being developed and to identify the weaknesses of the architectural design [146]. Several privacy threat analysis methodologies have been reported in the literature, to systematically elicit privacy requirements. These methodologies differ depending on the approach which they adopt to identify the privacy threats.

Some methodologies such as those presented in [147] and [146] propose systematic steps, starting with the documentation of the system under analysis, in the form of a data-flow diagram. The elements of the data-flow diagram are then mapped to privacy threat categories, namely: (i) linkability, identifiability, non-repudiation, detectability, disclosure of information, unawareness, and non-compliance (in [146], which cast the threat categories as the opposite of privacy goals); or (ii) linkability, unawareness and intervenability (in [147]). The privacy threats are later analyzed using threat trees as in [146] or attack trees as in [147]. Unlike [146], the methodology presented in [147] quantifies the overall privacy attack on a system, based on a quantitative evaluation (a risk-based quantification) of the individual attacks in the attack trees developed in the earlier steps.

Others, such as [148], propose a semi-automated problem-based privacy threat identification methodology. Automated privacy threat graphs are used based on the requirements of the designated system. Their proposed methodology is independent of specific privacy goals. Although their methodology is semi-automated, they use high-level privacy requirements and they do not provide detailed privacy knowledge for detailed analysis, unlike [146] and [147].

Other methodologies, such as the one presented in [149], view privacy requirements as an organizational goal and analyze where these goals are best implemented in the system. A mapping, between privacy requirements and the related privacy techniques, is used to determine the specific privacy enhancing techniques which will be adopted.

To the best of our knowledge, no research has to-date tried to identify the most important privacy services which must be included in WSN-based healthcare systems, with a view to improve the level of trust in such systems by patients, policy makers and regulators, and hence ease the acceptance of these systems by users and other stakeholders in real-world applications. It can be envisaged that to elicit the most important privacy services for a WSN-based healthcare system, a privacy threat methodology would be adopted to discover the list of such services for a given scenario. The methodologies reported in [146] and [147] are general-purpose systematic methods which could be adopted for the task.

## 5.2 Privacy assessment

In general, measuring the level of privacy afforded by a healthcare system is highly important to allow the assessment or comparison among different system designs, or to determine what needs to be improved in an existing system, and what impact this improvement would have on the different characteristics of the system, such as reliability, usability and privacy protection [150]. Where possible, a privacy assessment metric should be applied to determine the level of privacy achieved or achievable by privacy protection services.

This section discusses metrics used to gauge the achievable level of privacy. To the best of our knowledge, there have not been special measurement techniques or metrics specifically targeted at wireless sensor networks. Thus, the metrics discussed in this section are general ones. The main idea of these measurements is to assess the vulnerability to privacy-related attacks.

### 5.2.1 Metrics for privacy protection

Several metrics have been adopted in the literature to measure the level of anonymity, such as the anonymity set size and entropy [150]. The anonymity set refers to the number of senders and receivers of the messages within the network. The larger the anonymity set, the higher the level of the anonymity [151]. In addition, an ideal anonymous network should have an equal probability distribution for messages over the set of senders and receivers. This ideal is difficult to achieve in real life, and observation of the traffic flow over time by adversaries could allow them to narrow down the anonymity set and the probabilities associated with the senders or receivers [151].

The use of entropy was proposed by [152] and [153] to quantify the degree of anonymity. Entropy has been shown to be a useful way to quantify the level of anonymity for systems where the anonymity-set metric is not accurate enough to use [61] [150]. In relation to sender anonymity, the entropy of an anonymity set $S$ can be defined as $H(S) = -\sum_{i=1}^{N} p_i log_2(p_i)$, where $p_i$ is the probability of message $i$ being the actual sender of a particular message among $N$ other users [151].

Although entropy is said to be a better measure for anonymity than the anonymity set, several issues should be considered when using the entropy metric. Shannon's entropy estimates an average, and thus it might not provide a good insight into the worst case scenario of the anonymity measure. The use of other entropy measures such as min-entropy might give a better insight. In addition, for the entropy measure to be useful, one should learn all the possible distributions of the actions and actors, and the information that the adversary knows about the system or learns by observation. All this information is very hard to gather, especially in complex systems. Furthermore, the communication channels are intended to be used over the long term. However, entropy measures are often based on a rather limited number of interactions among specific actors and actions; as such, entropy measures might not provide a robust assessment for the system under consideration [150].

Furthermore, Syverson [154] warns that the conception of anonymity as indistinguishability within a set (of possible senders or receivers of a message, for example) — what he calls "the entropist conception of anonymous communication" — is not appropriate for general communication on large diversely shared networks (the Internet being one example). He argues that although entropy could say something about the uncertainty relating to who are senders or receivers of messages, entropism does not tell much about the amount of knowledge (or lack of knowledge) possessed by an adversary, and has led to system assumptions and adversary models which are not reasonable in practice.

In the same vein as the entropy of a message source has been used to quantify anonymity, it has been suggested that unlinkability can also be quantified using either probabilities or entropies [17].

Location privacy can assessed using a measure of information leakage (i.e. privacy loss) which is derived from the entropy equation [155]. Analysis of information leakage can be categorized according to the type of traffic analysis attack: correlation-based source identification attack, routing trace back attack, and reducing source space attack [156].

### 5.2.2 Assessment of the technical and economic costs

In addition to the measuring the level of privacy in its own right, other factors should be considered when assessing a healthcare system under construction. It is important to also estimate the technical and economic costs of including the privacy services in a healthcare system, when assessing the feasibility of these systems [155].

## 6 REGULATORY FRAMEWORKS FOR HEALTHCARE PRIVACY AND IMPLICATIONS FOR PRIVACY PROTECTION TECHNOLOGY

Privacy preservation is a complex issue in healthcare systems which are based on wireless sensor networks (see Section 2.4.2). From a regulatory point of view, some of the key factors are: the continuous capturing of medical data for long periods of time; the diverse and wide range of data about the medical and daily routines of patients or other subjects; and the potential use of the given health information in different applications by a wide range of beneficiaries such as insurance companies, life coaches, family, homecare providers, researchers and others [157]. Furthermore, there is the possibility of medical identity theft, for example in situations where employees who have access to patients records might sell this classified information to third parties, or the identity of a patient might be forged by another person to receive medication they are not entitled to [158]. Hence, laws to protect the privacy of healthcare subjects have been passed in many countries around the world. Although significant technological advances have been made towards effective healthcare systems based on wireless sensor networks, there is a need for a systematic evaluation of how well they meet privacy laws and other regulatory frameworks relating to the privacy of the health information of patients or other subjects [158].

Privacy of personal information is a legally protected right in many countries. However, different countries impose different laws which provide the legal foundations and regulatory frameworks or programs for healthcare privacy. For example, in the United States of America, the Health Insurance Portability and Accountability Act (HIPAA) was passed in 1996, in relation to privacy rights and policies for health information (including the protection of health information in electronic form) in the U.S., particularly with regards to portability and accountability across healthcare insurers and providers [159]. Another act, the American Recovery and Reinvestment Act (ARRA), was passed in 2009; it addresses some of the weaknesses of HIPAA [159]. Most countries in Europe have data-protection and privacy laws which also cover the privacy of health information. With regards to electronic health records, privacy practices of some member states of the European Union, such as the UK, are based on the EU Data Protection Directive, which came into force in 1998, to regulate the handling (automated or otherwise) of personal data [159].

Governments seek to constantly enhance the legal and other regulatory frameworks for healthcare, to guarantee the privacy rights of citizens. However, no direct mapping has been made between those frameworks and the privacy services which have been developed by technologists. It is however well understood that healthcare workflows and information systems must comply with the relevant regulations. A mapping between the functional requirements of technical solutions and legal principles is thus needed, to ensure that healthcare information systems will meet legal and other regulatory requirements. A good foundation towards the mapping was laid by Avancha et al. [159] (see TABLE IV. ) who elicited a set of actionable legal principles to provide the grounding for a conceptual privacy framework which specifies privacy properties to be provided by privacy-aware mobile-health information systems.

Avancha et al. [159] advocate that information systems for healthcare be grounded in fundamental principles which govern the protection of the privacy of patients. They thus defined a conceptual privacy framework, as a set of actionable principles obtained after distilling essential principles from four frameworks produced in the United States, respectively by the Office of the National Coordinator, the Health Privacy Project, the Markle Foundation, and the Certification Commission for Healthcare Information Technology. TABLE IV. shows the principles of the conceptual privacy framework and the corresponding set of privacy properties for mobile-health systems.

It should however be noted that the regulatory frameworks (including laws and policies), which govern individuals and organizations, vary across geographical and (sometimes) institutional boundaries and they also change over the years. For example, the Markle Foundation [160] launched the Health Common Framework for Private and Secure Health Information Exchange (Markle Common Framework). This framework provides the basis for secure, authorized, and private health information sharing between patients and authorized health providers. The framework was initially issued in 2008. In 2012, the Markle Foundation made amendments to the original framework to address the needs of contemporary healthcare systems. In the European Union, the EU Data Protection Directive [161] is an important directive for privacy protection. It states the conditions to be met for processing (collecting, storing, modifying, deleting, retrieving, transforming) personal data. The stated conditions relate to informed consent, legitimacy of purpose, and relevance of the data to the purpose.

TABLE IV.   LEGAL PRINCIPLES EMBEDDED IN THE CONCEPTUAL PRIVACY FRAMEWORK PROPOSED BY AVANCHA ET AL. [159] AND THE CORRESPONDING SET OF PRIVACY PROPERTIES TO BE PROVIDED BY A PRIVACY-AWARE MOBILE-HEALTH SYSTEM. (A) REFERS TO THE MAPPING BETWEEN THE LEGAL PRINCIPLES AND THE PRIVACY PROPERTIES. (B) IS THE INDEX FOR THE LABELS USED IN THE PRIVACY PROPERTIES MAPPING TABLE (A)

(A)

| Principle | Privacy properties of mHealth systems | | | | | | | | | | |
|---|---|---|---|---|---|---|---|---|---|---|---|
| | 1 | 2 | 3 | 4 | 5 | 6 | 7 | 8 | 9 | 10 | 11 |
| Openness and transparency | √ | √ | √ | | | | | | | | |
| Purpose specification | √ | | | | | √ | √ | | | | |
| Collection limitation and data minimization | | | √ | | | √ | √ | | | | |
| Use limitation (Transitive) | | | | | | | √ | | | | |
| Individual participation and control | | √ | √ | | | | | | | | |
| Data quality and integrity | | | | √ | | | | √ | | | |
| Security safeguards and controls | | | | | | | | | | | |
| Accountability and remedies | | | | | | | | | | √ | √ |
| Patient access to data | | | | √ | | | | | | | |
| Anonymity of presence | | | | | | | | | √ | | |

(B)

| Labels used for privacy properties of mHealth systems in table IV | |
|---|---|
| Index | Privacy properties |
| 1 | Inform Patients |
| 2 | Enable Patients to review storage and use of their PHI |
| 3 | Enable Patients to control, through informed consent |
| 4 | Provide access to PHI |
| 5 | Provide easy-to-use interfaces for all of the above |
| 6 | Limit collection and storage of PHI |
| 7 | Limit use and disclosure of PHI to those purposes previously specified and consented |
| 8 | Ensure quality of PHI |
| 9 | Hide Patient identity, sensor presence and data-collection activity from unauthorized observers |
| 10 | Support accountability through robust mechanisms |
| 11 | Support mechanisms to remedy effects of security breaches or privacy violations |

Further analysis and studies should be conducted to map the applicable regulatory frameworks for privacy protection onto explicit technical requirements. An example of work in this direction is given in [162] which offers a framework to extract privacy and security requirements from laws and regulations. Mapping regulatory frameworks onto technical requirements will allow the researchers and practitioners who develop healthcare systems to make sure that the systems comply with legal and other frameworks, and they would be acceptable (from a regulatory perspective) by patients, governments and other stakeholders. Failure to comply with the law, risks compromising the acceptance of the healthcare system even though it might be technically advanced and working efficiently.

In addition, Winkler and Rinner [132] advocate for what they call "controlled flexibility", whereby the requirements for the privacy and security protection techniques should depend on the application context (such as the legislative setting and cultural differences), and hence adaptability of the requirements is needed.

# 7   OPEN CHALLENGES AND FUTURE RESEARCH DIRECTIONS

The survey of the literature, on privacy in healthcare systems, which are based on wireless sensor networks, has revealed that there is a need for a lot more research specifically targeted at privacy, either alone or in synergy with security. The survey has uncovered many knowledge gaps, which require the attention of researchers and practitioners. They include gaps relating to the following challenges:

- Safeguarding both privacy and security through synergistic solutions
- Privacy violations perpetrated through intelligent computing systems or tools
- Adaptive privacy enhancing technology based on intelligent analytics
- Scalability of privacy enhancing technology
- Systematic and rigorous assessment of privacy enhancing technology
- Socio-technical systems which address privacy vulnerabilities emanating from humans and machines

An explanation of these challenges is as follows:

- *Safeguarding both privacy and security through synergistic solutions*: The literature is replete with papers wherein the distinction between technical solutions for privacy and security is rather blurred, and no attempt is made in those papers to systematically investigate the complementarity and trade-offs between security and privacy protection measures, for example with respect to access control and audit. Research into a clearer demarcation between the two aspects is required, possibly beginning from a clear definition of privacy goals and of their security counterparts, to be met by healthcare systems. The clear definition of privacy and security goals could then underpin the development of synergistic solutions, which exploit the complementarity between security and privacy protection measures.
- *Privacy violations perpetrated through intelligent computing systems or tools*: With more and more intelligence being built into sensing and communication devices [163], the world is moving towards a future where smart computing systems or tools, including wireless sensor networks, will be capable of operating autonomously to a significant extent, possibly with the help of data analytics applied to the so-called big data, such as through cloud-based services, for example. It can be envisaged that intelligent computing systems or tools endowed with semantic understanding capability could extract meanings from private information gathered by wireless sensor networks. Such intelligent computing devices attacking from outside a wireless sensor network, or smart malicious nodes attacking within the wireless sensor network, could be capable of circumventing data obfuscation techniques or other privacy protection techniques, by (for instance) deploying reasoning techniques which can infer the missing or hidden information by combining data from multiple sources. They may also be able to learn and change, such that they can adapt to new countermeasures deployed to combat them. Current countermeasures, which operate in a fairly pre-scripted manner, will not be able to tackle such malicious intelligent devices.

These devices would also raise many legal and ethical fears. For example, apportioning liability for privacy breaches could be difficult, given that the level human intervention in an actual attack could be low or non-existent. Technical and legal advances to address such concerns are required. For example, it can be envisaged that research into adaptive privacy protection countermeasures also endowed with intelligence and a high degree of autonomy will be required. These research efforts could be conducted in a symbiotic and cooperative relationship with work in the security area, such as machine learning and data mining methods for analytics in support of intrusion detection [164], or adversarial machine learning [165][166].

- *Adaptive privacy enhancing technology based on intelligent analytics*: Related to challenges discussed above, is the challenge of producing automated analysis techniques which can assess systematically and comprehensively the vulnerabilities of a healthcare system, such as by learning and making inferences about threats and attacks, and which can autonomously or semi-autonomously deploy appropriate preventative countermeasures or alerts.

- *Scalability of privacy enhancing technology*: Privacy protection in healthcare is a requirement, which permeates across a wide range of scales, from a single wireless sensor node through to a wireless sensor network which is integrated into a large-scale healthcare system. The latter is often required to handle high volumes of data (together with all the privacy implications of multimedia data and of the so-called 'big data'), or it may span wide geographical areas (within or across urban, national or continental boundaries), through platforms such as the Internet of Things (IoT) or cloud computing, for example. Such platforms are afflicted with vulnerabilities, which can be exploited by attackers, to violate privacy. A move from a tightly coupled closed architecture to integrated open systems is likely to introduce loopholes, such that vulnerabilities of one system may have a significant impact on the vulnerability of the whole system. A possible example scenario in the real world could be the integration of a healthcare wireless sensor network into a home IoT system, which is itself connected to the Internet.

A monolithic approach to privacy protection in systems such as those described above is unlikely to be successful. Instead, it is likely that the solutions would require a seamless integration of several privacy protection techniques, each tuned to specific operational conditions or circumstances (for example, subsystems may each have their own privacy policies and requirements which may differ from those of other subsystems). Self-tuning to specific operational conditions or circumstances would be beneficial, and it could be achieved through built-in context awareness, to balance privacy and data utility (for example).

The development of the solutions discussed above should also include a consideration of the optimality of the privacy preservation techniques. Optimality could, for example, be defined with respect to the storage, processing, and communication requirements of wireless sensor networks; this is particularly important if multimedia sensors are also included, or the network is a wireless body sensor network. Optimality could also be specified in relation to the demand for rapid access to healthcare data in an emergency, whereby privacy restrictions may need to be overridden or adjusted as dictated by circumstances.

Scalability would also need to satisfy the requirement that the integrated system complies with the applicable workflow, and policy or regulatory schemes across institutional boundaries or geographical borders. There will also be a requirement that the integration mechanisms include checks for consistency and resolution of conflicts between different schemes.

- *Systematic and rigorous assessment of privacy enhancing technology*: There is an urgent need for methods for conducting a systematic and rigorous assessment of the ability of a healthcare system, which includes wireless sensor networks, to withstand attacks injurious to privacy. Such an assessment requires methods, which can detect and find vulnerabilities that can be exploited by attackers, including metrics for measuring the level of protection against privacy breaches. Quantitative and qualitative metrics are required for a rigorous assessment, which is appropriate for healthcare systems already deployed in the real world or those being designed for such deployment. In particular, given the special constraints and operational environments of wireless sensor networks, an investigation into metrics specifically targeted at such networks in healthcare contexts would be worthwhile.

Performing a privacy assessment of a system faces a number of challenges. One important challenge is the dearth of established objective metrics for measuring reliably the degree of protection afforded by an attack countermeasure. Another challenge is the difficulty of assessing with confidence the impact, which vulnerabilities that reside inside individual system components can have on the privacy properties of a system, which integrates those components. The legal implications of this challenge may be exacerbated when the individual components of the system are in different geographical areas (possibly located in different countries, which may have different regulatory requirements) and are operated by different stakeholders along the healthcare provision chain (possibly with different operational policies to be enforced along the chain). In addition, the impact of vulnerabilities associated with the people who use the system is another important factor that is difficult to assess.

More research is thus needed into methodologies for privacy assessment, which include the evaluation of privacy protection at different levels-of detail (from the individual components up to the whole system which integrates the components), using appropriate quantitative or qualitative metrics, and taking into account all key factors, including the humans who use the system. Viewing the wireless sensor network as a communication network, the level-of-detail analysis could include a mapping of attacks and countermeasures to each layer of the network communication protocol, given that wireless communication is fundamental to the operation of wireless sensor networks. For instance, work on such mappings could yield a generic layered architecture for healthcare systems based on wireless sensor networks, including proposals or specifications of the services embedded in each layer or hosted on each component (such as the sensor, the gateway, or the end-system). This architecture could be linked to a mapping between the privacy services and the Open Systems Interconnection model (OSI model), for example.

The work on privacy assessment methodologies could build upon the basic dimensions (namely: impact, target, source, and vulnerability) which were identified by Igure and Williams [22] for classifying attacks in computer systems. Other possible further work could focus on applying privacy threat analysis and risk assessment methodologies to healthcare systems, which are based on wireless sensor networks, to produce a priority list of privacy and security services required for such systems. The research could extend the already published work on creating attack trees or threat trees [146] [147], to document the details of possible attacks, for further analysis and consideration in existing systems or during the design of systems under construction.

- *Socio-technical systems which address privacy vulnerabilities emanating from humans and machines*: Human vulnerabilities are often a point-of-failure targeted by attackers. Hence, there is an urgent need for research into privacy protection, which goes beyond mere technological considerations. This will require a holistic interdisciplinary approach which covers research and practice areas from various fields such as computing, medicine, psychology, law, …

From a regulatory point of view, for example, it will be important that legal requirements are codified into operational requirements to be fulfilled by privacy enhancing techniques. The codification could involve establishing a legal and technical mapping between the legal frameworks for healthcare and the privacy services embedded into computer-based privacy protection systems. The laws for healthcare privacy which are discussed briefly in Section 6 provide actionable principles to protect healthcare privacy [157]. Healthcare systems based on wireless sensor networks must comply with privacy laws and policies issued by the governments and regulators.

A possible approach to ensure this compliance is to directly map the regulatory frameworks to the privacy services; an illustrative instance of this approach is given in [162]. For example, in order for consumers to be confident that their data is collected and accessed in line with privacy legislation, assurances must be given for the privacy afforded by privacy protection services with respect to data in transit, data queries, and data delivery. Furthermore, there is need for robust privacy and security safeguards and controls, which address the applicable regulatory requirements.

## 8   CONCLUSIONS

This paper presents a literature survey on privacy protection techniques in wireless sensor networks for healthcare, including those, which use multimedia sensors. It highlights sample applications, and includes a discussion of privacy issues and the imperative need for privacy protection in healthcare, and it spells out the multifaceted challenge of privacy protection in wireless sensor networks for healthcare. To avoid mixing privacy and security concerns, as is often the case in the literature, the survey untangles them by decoupling privacy goals from security goals. These goals then guide the review of privacy protection in wireless sensor networks, with a focus on privacy services and their underpinning mechanisms to achieve privacy goals or to protect location privacy. In addition, included in this part of the review are developments towards privacy protection in healthcare systems, which are built on wireless multimedia sensor networks. In addition, the paper reviews methods for performing threat analysis and for assessing the level of privacy protection afforded by a system. Although the main thrust of the paper is on hard privacy, it also briefly enters the realm of soft privacy by considering the regulatory frameworks for privacy in healthcare and their implications for privacy protection technology. The paper winds up with a discussion of open research challenges, and a proposal of avenues for future research.

Wireless sensor networks exploit the availability of miniature wearable biosensors, which have generated new opportunities in many sectors of life in the twenty-first century, such as the deployment of wireless sensor networks in healthcare. For many potential applications in the healthcare sector, the hardware for capturing, storing, and processing the relevant data using wireless sensor networks is readily available at reasonably low costs. Moreover, significant advances in data processing techniques and wireless communications already allow some useful healthcare applications, if privacy and security requirements can be met.

The uses of wireless sensor networks in healthcare applications are motivated by the need to provide quality healthcare, to treat or (ideally) prevent health problems. Modern computing tools can make a significant contribution towards meeting this need. For instance, there is a demand for diagnostic and health monitoring tools, which would operate everywhere and all the time. In particular, wireless sensor networks can be a key component for assistive technology to enable the elderly or people with disabilities or with chronic health problems to live as independently as possible, and for tools to monitor human activity and vital signs in neonatal healthcare or in general wellbeing programs, and for many other aspects of healthcare in the twenty-first century. However, the acceptance of wireless sensor networks by healthcare stakeholders is influenced by the effectiveness of the techniques deployed to safeguard the personal (and possibly intimate) information which is captured and transmitted over the air, within and beyond the wireless sensor networks. This is currently an important stumbling block on the way to widespread deployment of wireless sensor networks in the healthcare domain. Privacy preservation is thus an important consideration in the development and deployment of wireless sensor networks for healthcare applications. However, despite its importance, fewer research papers address privacy preservation (compared to security) in wireless sensor networks for healthcare. Significant technical and socio-technical advancements are needed.

Despite some advances, research into privacy protection for wireless sensor networks in healthcare is not sufficiently mature. Over the years, privacy preservation for wireless sensor networks has evolved from being an off-shoot of security protection, into a full-fledged research concern in its own right. However, techniques which safeguard the privacy of the data subject and other healthcare stakeholders, over a wide range of operational conditions and in compliance with the applicable regulations and policies, are still being sought.

Further research is required on many fronts, including: methods for systematic and rigorous assessment of privacy enhancing technology; socio-technical systems which address privacy vulnerabilities emanating from humans and machines; scalable privacy enhancing technology; synergistic solutions for safeguarding both privacy and security; technical and legal advances to address privacy violation emanating from intelligent computing systems or tools; adaptive privacy enhancing technology based on intelligent analytics. Progress in these areas will unlock the door onto effective and efficient privacy protection, which is appropriate for healthcare applications, and will allow wireless sensor networks to play a major role in healthcare provision fit for the twenty-first century.

## APPENDIX A: DEFINITIONS

*Accountability*: "Property that ensures that the actions of an entity may be traced uniquely to the entity". [167]
*Anonymity*: "Anonymity of a subject means that the subject is not identifiable within a set of subjects, the anonymity set". [168]
*Attack*: That which realizes a threat (i.e. an attack actually exploits a vulnerability), and is carried out by an *attacker*. In healthcare, the attacker could be the patient, authorized personnel, or a third party [25].
*Authenticity*: "Property that an entity is what it claims to be". [169]
*CIA triad*: Three properties (Confidentiality, Integrity, Availability) required of security systems. They include:
• *Confidentiality*: "Property that information is not made available or disclosed to unauthorized individuals, entities, or processes". [169]
• *Integrity*: "Property of accuracy and completeness". [169]
• *Availability*: "Property of being accessible and usable upon demand by an authorized entity". [169]
*Countermeasure*: That which can mitigate a vulnerability.
*Electronic Health Record*: Medical or other health records created and managed by a healthcare provider.
*Healthcare*: "The prevention, treatment, and management of illness and the preservation of mental and physical well-being through the services offered by the medical and allied health professions." [170]
*Healthcare provider*: Organization which provides healthcare services – such as hospitals and nursing homes for the elderly, for example.
*Health information privacy*: "An individual's right to control the acquisition, uses, or disclosures of his or her identifiable health data." [171]
*Information security*: "Preservation of confidentiality, integrity and availability of information". [169]
Note: "In addition, other properties, such as authenticity, accountability, non-repudiation, and reliability can also be involved". [169]
A much more explicit definition of information security is provided in [172] as "The term 'information security' means protecting information and information systems from unauthorized access, use, disclosure, disruption, modification, or destruction in order to provide—
(A) integrity, which means guarding against improper information modification or destruction, and includes ensuring information nonrepudiation and authenticity;
(B) confidentiality, which means preserving authorized restrictions on access and disclosure, including means for protecting personal privacy and proprietary information; and
(C) availability, which means ensuring timely and reliable access to and use of information."
*Non-repudiation*: "Ability to prove the occurrence of a claimed event or action and its originating entities". [169]
*Personal Health Record*: Medical or other health records created and managed by the subject (a patient, for example).
*Personal Health Information*: "Any personally identifiable information (PII) that is health-related. We follow NCVHS and define PHI as "personal health information" rather than "protected health information", which is a phrase that has specific meaning in a specific legal context (HIPAA) [171]. HIPAA defines "protected health information" as individually identifiable health information that is stored or transmitted in any medium, including information related to the Patient's health, the provision or healthcare, or billing for healthcare.

Contrast with *PII* and *PrHI*." [159]

*Personally Identifiable Information*: "Any information that uniquely identifies an individual, such as a name, social security number, patient ID number, or street address." [159]

*Privacy service*: A software component, which provides privacy-enhancing functionality to other components of a healthcare system. A service may be invoked on its own or combined with other services to provide privacy protection.

*Privacy mechanism*: A processing technique (or a device which incorporates such a technique) which handles privacy attacks.

*Pseudonymity*: "Pseudonymity is the use of pseudonyms as identifiers", instead of using a real name of the subject. [168].

*Reliability*: "Property of consistent intended behavior and results". [169]

*Threat*: That which can potentially exploit a vulnerability.

*Unlinkability*: "Unlinkability of two or more items of interest (IOIs, e.g., subjects, messages, actions, ...) from an attacker's perspective means that within the system (comprising these and possibly other items), the attacker cannot sufficiently distinguish whether these IOIs are related or not" [168].

*Undetectability*: "Undetectability of an item of interest (IOI) from an attacker's perspective means that the attacker cannot sufficiently distinguish whether it exists or not" [168].

*Unobservability*: "Unobservability of an item of interest (IOI) means

- undetectability of the IOI against all subjects uninvolved in it, and
- anonymity of the subject(s) involved in the IOI even against the other subject(s) involved in that IOI". [168]

*Vulnerability*: That which can cause a violation of privacy.

APPENDIX B: LIST OF ACRONYMS

| | |
|---|---|
| BAN | Body Area Network |
| EHR | Electronic Health Record |
| IEC | International Electrotechnical Commission |
| IETF | Internet Engineering Task Force |
| IoT | Internet of Things |
| ISO | International Organization for Standardization |
| ITU | International Telecommunication Union |
| MEMS | Micro-Electromechanical Systems |
| M2M | Machine-to-machine communication |
| PAN | Personal Area Network |
| PET | Privacy Enhancing Technology |
| PHR | Personal Health Record |
| PHI | Personal Health Information |
| PII | Personally Identifiable Information |
| RFID | Radio-Frequency Identification |
| SoC | System on a Chip |
| WSN | Wireless Sensor Network |
| WMSN | Wireless Multimedia Sensor Network |
| WBSN | Wireless Body Sensor Network |